\documentclass[epj]{svjour} 
\usepackage{iftex}

\ifPDFTeX
  \usepackage{ucs}
  \usepackage[utf8]{inputenc}
  \usepackage[T1]{fontenc}
  \usepackage[english]{babel}
\else
  \ifXeTeX
    \usepackage{xltxtra}
    \usepackage{polyglossia}
    \setmainlanguage[variant=british]{english}
    \setotherlanguage[spelling=new, latesthyphen=true]{german}
    \usepackage{fontspec}
    \defaultfontfeatures{Ligatures=TeX}
    \defaultfontfeatures{Mapping=tex-text}
  \else
    \usepackage{luatextra}
  \fi
\fi

\usepackage{amsmath,amssymb}
\usepackage{graphicx}
\usepackage{xcolor}
\usepackage{subfigure}
\usepackage{hyperref}
\usepackage{csquotes}

\usepackage{mathtools}

\usepackage{todonotes}
\usepackage{nicefrac}

\usepackage{hepnames}
\usepackage{physics}

\usepackage{bm}
\usepackage{siunitx}

\usepackage{todonotes}

\newcommand{\fig}[1]{Figure~\ref{#1}}
\newcommand{\tab}[1]{Table~\ref{#1}}
\newcommand{\eq}[1]{Eq.~(\ref{#1})}

\DeclarePairedDelimiterX\set[1]\lbrace\rbrace{#1}

\newcommand{\SO}{	\ensuremath{\mathrm{SO}} }

\newcommand{\G}{	\ensuremath{\mathit{G}} } 
\newcommand{\LG}{	\ensuremath{\mathrm{LG}} } 

\newcommand{\Oh}{		\ensuremath{\mathrm{O}_\text{h}} } 	
\newcommand{\Cfourv}{	\ensuremath{\mathrm{C}_\text{4v}} } 
\newcommand{\Ctwov}{	\ensuremath{\mathrm{C}_\text{2v}} } 
\newcommand{\Cthreev}{	\ensuremath{\mathrm{C}_\text{3v}} }

\newcommand{\R}{\ensuremath{\hat{R}}\xspace} 

\newcommand{\pcm}{\bm{p}_\text{cm}\xspace}

\newcommand{\eto}[1]{\ensuremath{\mathrm{e}}^{#1}}
\renewcommand{\i}{\mathrm{i}}

\newcommand{\sommer}{\ensuremath{r_0}\xspace}

\begin{document}

\title{\boldmath Hadron-Hadron Interactions from $N_f=2+1+1$ Lattice QCD:\\
  The $\rho\,$-resonance}

\newcommand{\bn}{HISKP and BCTP, Rheinische Friedrich-Wilhelms-Universität, Bonn, Germany}

\author{M.~Werner\inst{1}
  \and M.~Ueding\inst{1}
  \and C.~Helmes\inst{1}
  \and C.~Jost\inst{1}
  \and B.~Knippschild\inst{1}
  \and B.~Kostrzewa\inst{1}
  \and C.~Liu\inst{2,3}
  \and L.~Liu\inst{4}
  \and B.~Metsch\inst{1}
  \and M.~Petschlies\inst{1}
  \and C.~Urbach\inst{1}
}
\institute{
  Helmholtz-Institut~f\"{u}r~Strahlen-~und~Kernphysik~and~Bethe~Center~for
  Theoretical~Physics, Universit\"{a}t~Bonn, \\ Bonn,~Germany
  \and School of Physics and Center for High Energy Physics,                                                   
  Peking University, \\ Beijing,~China
  \and Collaborative Innovation Center of Quantum Matter, \\ Beijing,~China
  \and Institute of Modern Physics, Chinese Academy of Sciences, \\ Lanzhou,~China
}

\date{Received: date / Revised version: date}

\abstract{
  We present a lattice QCD investigation of the $\rho$-meson with
  $N_f=2+1+1$ dynamical quark flavours for the first time. The
  calculation is performed based on gauge 
  configuration ensembles produced by the ETM collaboration with three
  lattice spacing values and pion masses ranging from
  $230\ \mathrm{MeV}$ to $500\ \mathrm{MeV}$. Applying the Lüscher
  method phase shift curves are determined for all ensembles
  separately. Assuming a Breit-Wigner form, the $\rho$-meson mass and width are
  determined by a fit to these phase shift curves. Mass and width combined
  are then extrapolated to the chiral limit, while lattice artefacts
  are not detectable within our statistical uncertainties. For the
  $\rho$-meson mass extrapolated to the physical point we find good
  agreement with experiment. The corresponding decay width differs by 
  about two standard deviations from the experimental value.
  \PACS{
    {11.15.Ha}{}  \and
    {12.38.Gc}{} \and
    {12.38.Aw}{} \and     
    {12.38.-t}{} \and     
    {14.70.Dj}{}
  } 
}

\maketitle

\section{Introduction}

The $\rho$-meson represents together with the (in-)famous
$\sigma$-meson ($f_0(500)$) one of the most prominent meson resonances
in the standard model. The $\rho$ 
decays almost exclusively to two pions and the experimental phase
shift curve~\cite{Erwin:1961ny,Protopopescu:1973sh} is a textbook example for a relativistic
Breit-Wigner form. Moreover, the $\rho$, being the lightest vector meson,
plays a fundamental role in many processes within the context of vector meson
dominance, for a review see Ref.~\cite{Meissner:1987ge}. 

Therefore, an investigation of the $\rho$-meson properties from first
principles with lattice QCD is highly desirable. However, unstable
particles require special care in lattice QCD: interaction properties
can only be computed using the by now famous Lüscher
method~\cite{Luscher:1985dn,Luscher:1986pf,Luscher:1990ux}. With its
help, infinite volume scattering properties can be extracted from
finite volume energy shifts. In the meanwhile 
the Lüscher method has been developed further in many
directions, for a review see Ref.~\cite{Briceno:2017max}, in particular
also for three particle systems, see for instance
Refs.~\cite{Polejaeva:2012ut,Briceno:2018aml,Romero-Lopez:2018rcb,Pang:2019dfe,Hansen:2019nir,Mai:2017bge}. For
this paper most relevant is the derivation of 
the formalism in moving
frames~\cite{Rummukainen:1995vs,Feng:2010es,Gockeler:2012yj}, which
allows one to map out 
the phase shift at many different scattering momenta, without the need
to study different volumes. 

For a long time the Lüscher method was difficult to apply to the
$\rho$ in realistic lattice calculations, albeit there are some early 
attempts~\cite{McNeile:2002fh,Michael:2006hf}.
By now, there are a number of investigations of the $\rho$-meson from
lattice QCD using the Lüscher
method~\cite{Feng:2010es,Lang:2011mn,Aoki:2011yj,Dudek:2012xn,Bali:2015gji,Wilson:2015dqa,Fu:2016itp,Guo:2016zos,Alexandrou:2017mpi,Andersen:2018mau}. The 
first computation with light dynamical up and down quarks can be found
in the pioneering work of Ref.~\cite{Feng:2010es}. Subsequent
investigations focused on different aspects like large operator
bases~\cite{Dudek2013} or asymmetric boxes~\cite{Guo:2016zos}. Recently, a first
investigation involving different lattice spacings and a range of pion
masses has been performed~\cite{Andersen:2018mau}. However, in the latter
reference chiral and continuum extrapolations were not performed.

With this paper we fill this gap and present a computation of the
$\rho$-meson applying the Lüscher method using gauge ensembles
generated with $N_f=2+1+1$ dynamical quark flavours by the ETM
collaboration at three different lattice spacing values and a wide
range of pion masses~\cite{Baron:2010bv,Baron:2010th}. This
allows us to perform a chiral and continuum extrapolation
of the $\rho$-meson mass and width. Note that in
Ref.~\cite{Giusti:2018mdh} the mass and width of the $\rho$-meson has
been determined on the same gauge configurations, however, using an
inverse Lüscher approach based on the vector current only combined
with a parametrisation of the pion form factor.

This paper is organised as follows: after presenting the lattice
action and its parameters in section~\ref{sec:action}, we discuss our
methods in section~\ref{sec:method}. In section~\ref{sec:results} we
present our results, the main result being the continuum extrapolated
values of $M_\rho$ and $\Gamma_\rho$ at the physical pion mass value
reading
\[
M_\rho\ =\ 769(19)\ \mathrm{MeV}\,,\qquad\Gamma_\rho\ =\ 129(7)\ \mathrm{MeV}\,.
\]
In section~\ref{sec:discussion} we discuss our results and put them
into perspective, followed by a summary in
section~\ref{sec:summary}. More technical details can be found in the
appendix.

\section{Lattice Action}
\label{sec:action}

The lattice details for the investigation presented here are very
similar to those we used in our previous studies on hadron-hadron
interactions~\cite{Helmes:2015gla,Helmes:2017smr,Helmes:2018nug,Helmes:2019dpy}. 
We use $N_f=2+1+1$ flavour lattice QCD
ensembles generated by the ETM Collaboration, for which details can be
found in Refs.~\cite{Chiarappa:2006ae,Baron:2010th,Baron:2010bv}. 
The parameters relevant for this paper are compiled in
Table~\ref{tab:setup}: we give for each ensemble the inverse gauge coupling
$\beta=6/g_0^2$, the bare values for the quark mass parameters $\mu_\ell, \mu_\sigma$
and $\mu_\delta$, the lattice volume and the number of configurations on
which we estimated the relevant quantities. 

\begin{table*}
 \centering
 \begin{tabular*}{.9\textwidth}{@{\extracolsep{\fill}}lcccccc}
  \hline\hline
  ensemble & $\beta$ & $a\mu_\ell$ & $a\mu_\sigma$ & $a\mu_\delta$ &
  $(L/a)^3\times T/a$ & $N_\mathrm{conf}$  \\ 
  \hline\hline
		A30.32   & $1.90$ & $0.0030$ & $0.150$  & $0.190$  & $32^3\times64$ & $623$  \\
		A40.24   & $1.90$ & $0.0040$ & $0.150$  & $0.190$  & $24^3\times48$ & $997$  \\
		A40.32   & $1.90$ & $0.0040$ & $0.150$  & $0.190$  & $32^3\times64$ & $493$  \\
		A60.24   & $1.90$ & $0.0060$ & $0.150$  & $0.190$  & $24^3\times48$ & $618$  \\
		A80.24   & $1.90$ & $0.0080$ & $0.150$  & $0.190$  & $24^3\times48$ & $611$  \\
		A100.24  & $1.90$ & $0.0100$ & $0.150$  & $0.190$  & $24^3\times48$ & $307$  \\
		\hline
		B25.32   & $1.95$ & $0.0025$ & $0.135$  & $0.170$  & $32^3\times64$ & $197$ \\
		B35.32   & $1.95$ & $0.0035$ & $0.135$  & $0.170$  & $32^3\times64$ & $493$ \\
		B35.48   & $1.95$ & $0.0035$ & $0.135$  & $0.170$  & $48^3\times96$ & $265$ \\
		B55.32   & $1.95$ & $0.0055$ & $0.135$  & $0.170$  & $32^3\times64$ & $613$ \\
		\hline                                                                 
		D15.48 & $2.10$ & $0.0015$ & $0.120$ & $0.1385$ & $48^3\times96$ & $304$ \\
		D30.48 & $2.10$ & $0.0030$ & $0.120$ & $0.1385$ & $48^3\times96$ & $241$ \\
		D45.32sc & $2.10$ & $0.0045$ & $0.0937$ & $0.1077$ & $32^3\times64$ & $588$ \\
  \hline\hline

 \end{tabular*}
 \caption{The gauge ensembles used in this study. For the labeling of
   the ensembles we adopted the notation in
   Ref.~\cite{Baron:2010bv}. In addition to the relevant input
   parameters we give the lattice volume and  the number of evaluated
   configurations, $N_\mathrm{conf}$.}
 \label{tab:setup}
\end{table*}

The ensembles were generated using the
$N_f=2+1+1$ twisted mass fermion
action~\cite{Frezzotti:2003ni,Frezzotti:2003xj,Frezzotti:2004wz}. 
For orientation, the $\beta$-values \numlist{1.90;1.95;2.10} correspond to
lattice spacing values of $a\sim\,0.089\ \mathrm{fm},
0.082\ \mathrm{fm}$ and $a\sim\,0.062\ \mathrm{fm}$,
respectively, see also \tab{tab:r0values}. The
ensembles were generated at so-called maximal twist, which guarantees
automatic order $\order{a}$ improvement for almost all physical
quantities~\cite{Frezzotti:2003ni}. The corresponding lattice Dirac
operator in the light sector reads
\begin{equation}
  D^\mathrm{tm}_\ell\ =\ D_\mathrm{W} + m_\mathrm{cr} + i\mu_\ell\gamma_5\tau^3\,,
\end{equation}
with $D_\mathrm{W}$ the Wilson Dirac operator, $m_\mathrm{cr}$ the
Wilson quark mass tuned to its critical value, $\mu_\ell$ the bare
up/down quark mass parameter and $\tau^3$ the third Pauli
matrix acting in flavour space. The tuning of the Wilson quark mass to
its critical value is discussed in Ref.~\cite{Baron:2010bv}, where
also the Dirac operator for the strange/charm sector can be found,
which is not relevant for the remainder of this paper. The relation of
the bare parameters $\mu_\sigma$ and $\mu_\delta$ given in
\tab{tab:setup} are related to the renormalised strange and charm
quark masses as follows:
\[
m_{s,c}\ =\ \frac{1}{Z_P} \mu_\sigma \mp \frac{1}{Z_S} \mu_\delta\,.
\]
The bare strange and charm quark masses are kept constant for each
$\beta$-value. The renormalised strange quark mass values differ from
the physical one by up to 10\%, see
Ref.~\cite{Baron:2010bv,Baron:2010th,Carrasco:2014cwa} for details.
In our chiral and continuum extrapolation we treat the strange quark
mass as constant in spite of this deviation. In the gauge sector the
Iwasaki action is used~\cite{Iwasaki:1983ck,Iwasaki:1984cj}. 

The biggest disadvantage of Wilson twisted mass fermions at maximal
twist is the breaking of isospin symmetry. As a consequence, charged
and neutral pions are not mass degenerate, with the splitting in the
squared masses vanishing like $a^2$ towards the continuum limit. This
pion mass splitting is also about the only quantity where strong
effects of isospin splitting have been observed so
far~\cite{Herdoiza:2013sla}. 

We are going to study the decay $\rho^0\to\pi^+\pi^-$ in a
$p$-wave. In nature, there is no mixing with two neutral pions
possible. Even if there is reduced isospin symmetry (only $I_z$ is a
good quantum number) in the Wilson
twisted mass formulation at maximal twist, such mixing is still not
possible due to $C$-symmetry: $\rho^0$ is $C$-odd, while $\pi^0\pi^0$
is $C$-even. Likewise, non $p$-wave symmetric combinations of $\pi^+\pi^-$ are
$C$-even, while $p$-wave symmetric combinations of $\pi^+\pi^-$ are
$C$-even, for instance 
\[
\begin{split}
  O^{l=0,1} &= \pi^+(p_1) \pi^-(p_2) + (-1)^l \pi^-(p_1) \pi^+(p_2)\\
  C\, O^{l=0,1}\, C^{-1} &= \pi^-(p_1) \pi^+(p_2) + (-1)^l \pi^+(p_1)
  \pi^-(p_2) \\
  &= (-1)^l O^{l=0,1}\,,\\
\end{split}
\]
excluding also mixings with $I=2, I_z=0$ states. Moreover,
also a single $\pi^0$ is $C$-even and cannot mix.

However, due to missing isospin symmetry, there are fermionic
disconnected contributions to the $\rho^0$ lattice interpolating
operators. These can be shown, like for the neutral pion, to be purely
of $\mathcal{O}(a^2)$. Thus, we drop them from our calculation, as was
also done in Ref.~\cite{Feng:2010es}. Note that the neutral to charged
$\rho$-meson splitting was found to be
negligible~\cite{Michael:2007vn}. 

As a smearing and contraction scheme we employ the stochastic
Laplacian-Heaviside (sLapH) approach, described in
Ref.~\cite{Morningstar:2011ka}. 
Details of our sLapH parameter choices can be found in
Refs.~\cite{Helmes:2015gla,Helmes:2017smr}. 

\begin{table}
  \centering
  \begin{tabular*}{.5\linewidth}{@{\extracolsep{\fill}}lrr}
    \hline\hline
    $\beta$ &
    $a\ [\mathrm{fm}]$ & $r_0/a$ \\
    \hline\hline
    $1.90$  & $0.0885(36)$ & $5.31(8)$ \\
    $1.95$  & $0.0815(30)$ & $5.77(6)$ \\
    $2.10$  & $0.0619(18)$ & $7.60(8)$ \\
    \hline\hline
  \end{tabular*}
  \caption{Values of the Sommer parameter $r_0/a$ and the
    lattice spacing $a$ 
    at the three values of $\beta$. See 
    Ref.~\cite{Carrasco:2014cwa} for more details.}
  \label{tab:r0values}
\end{table}

\subsection{Scale Setting}

The scale setting for the ensembles used here has been performed in
Ref.~\cite{Carrasco:2014cwa} by extrapolating pseudo-scalar meson
masses and decay constants to the chiral and continuum limits and
using the physical values of $M_\pi$ and $f_\pi$ as inputs. As an
intermediate scale the Sommer parameter $r_0/a$ has been used. The
values for the lattice spacings resulting from this procedure can be
found in \tab{tab:r0values} together with the values of $r_0/a$
for each $\beta$-value. The physical value of the Sommer parameter was
determined in Ref.~\cite{Carrasco:2014cwa} on the same ensembles as
the value 
\begin{equation}
  r_0 = 0.474(11)\ \mathrm{fm}\,.
\end{equation}
In this paper we are also going to use the Sommer parameter as
intermediate lattice scale. In addition to the physical value for
$r_0$ given above we need the physical pion mass value as
input. Here, we use the value of $M_\pi$ in the isospin symmetric
limit~\cite{Aoki:2016frl} (consistent with what was used in
Ref.~\cite{Carrasco:2014cwa}) 
\begin{equation}
  \overline{M}_{\pi^+}\ =\ 134.8(3)\ \mathrm{MeV}
\end{equation}
corrected for QED and strong isospin contributions.
The values of $r_0/a$ were not determined by us on the identical set
of gauge configurations. Therefore, we use the values given in
\tab{tab:r0values} with re-sampling (parametric bootstrap). 
$\overline{M}_\pi$ and its error are treated in the same way.

In the appendix~\ref{sec:details} we discuss how we include
uncertainties on $r_0/a$, $\overline{M}_{\pi^+}$ and other input.
We remark that at fixed $\beta$-value there is in principle
correlation between $r_0/a$ and all other observables.
However, we cannot take these correlations
into account, because $r0/a$ was not determined on the identical gauge
configurations. However, we measured this correlation to more
precisely estimated quantities like $aM_\pi$ previously and found the
correlation to be negligible. 
 
\section{Methods}
\label{sec:method}

In this section we summarise the methodology we applied to extract our
results. 

\subsection{Scattering in Finite Volume}
\label{sec:luescher}

As is well known, the extraction of scattering properties from lattice
QCD in Euclidean space-time and a finite volume requires the
application of the so-called Lüscher method~\cite{Luscher:1986pf,Luscher:1990ux}. 
It allows one to relate finite volume induced energy shifts to infinite
volume scattering properties of $n$-particle systems in the continuum. 
The formalism is based on the following determinant equation
\begin{equation}
  \label{eq:luescher_formula}
  \det \left(M_{lm,l'm'}(k) - \delta_{ll'} \delta_{mm'} \cot(\delta_l) \right) = 0\,,
\end{equation}
where $M_{lm,l'm'}$ is an analytically known matrix function of the
lattice scattering momentum $k$, see below. $\delta_l$ is the phase
shift of the $l$-th partial wave and the determinant
acts in angular momentum space. In the case of pion-pion scattering
the lattice scattering momentum $k$ is related to a given energy value
$E_\mathrm{CM}$ in the centre-of-mass (CM) frame and the pion mass
$M_\pi$ via 
\begin{equation}
  \label{eq:scatteringk}
  k^2\ =\ \frac{E_\mathrm{CM}^2}{4} - M_\pi^2\,.
\end{equation}
Given the scattering momentum on the lattice, \eq{eq:luescher_formula}
thus yields $\delta_l$. In order to map out the dependence of
$\delta_l$ on $E_\mathrm{CM}$, as many values of
$E_\mathrm{CM}$ as possible must be extracted from a lattice calculation.

This is most conveniently done by using several CM momenta, as first
proposed in Ref.~\cite{Rummukainen:1995vs}. For given CM momentum $\pcm$,
the relativistic energy reads
\begin{equation}
  \label{eq:disp}
  W_\text{L} = \sqrt{\pcm^2 + E_\text{CM}^2}
\end{equation}
where $\pcm$ is, due to the finite volume, quantised as
\begin{align*}
  \pcm = \frac{2 \pi}{L} \cdot \bm{d} \,, \quad \bm{d} \in \mathbb{Z}^3 \,.
\end{align*}
We classify momentum sectors by $\abs{\bm{d}}^2$ and use all allowed
lattice momenta in each sector up to $\mathbf{d}^2 = 4$. We denote the
set of equivalent momenta as
\begin{equation*}
  \set{\bm{d}} \equiv \set{\bm{z} \in \mathbb{Z}^3 \,, \quad {\bm{z}}^2 = \bm{d}^2}\,.
\end{equation*}
By applying a corresponding Lorentz boost
\[
\gamma\ =\ \frac{W_\text{L}}{E_\mathrm{CM}}\,,
\]
we can compute $E_\mathrm{CM}$ for given $W_L$ and $\pcm$.
Adopting the notation of Refs.~\cite{Bernard:2008ax,Gockeler:2012yj}, it
remains to give details for the matrix $M$ from
\eq{eq:luescher_formula}. Its matrix elements are given by
\begin{equation}
  M_{lm,l'm'} = (-1)^l \sum_{j=\abs{l-l'}}^{l+l'} \sum_{s=-j}^{j} \sqrt{2j+1} \, i^j w_{js} C_{lm,js,l'm'}\,,
  \label{eq:luescher_matrix_lmlm}
\end{equation}
with the convenient notation
\begin{align}
  w_{js} = \frac{\mathcal{Z}_{js}(1, q^2)}{\pi^{3/2} \sqrt{2j+1} \gamma q^{j+1}} \,,\qquad q = \frac{k L}{2 \pi} \,.
\end{align}
$C_{lm,js,l'm'}$ represent coefficients which can be expressed using
Wigner $3j$-symbols, see Ref.~\cite{Gockeler:2012yj}.

In a finite volume the symmetry group of rotoflections (rotations and
space inversions) is reduced from $\mathrm{O}(3)$ to a finite
subgroup.\footnote{We treat parity explicitly instead of just looking
  at $\mathrm{SO}(3)$ because parity will not be conserved
  in moving reference frames.} Because $\pcm$ is an invariant,
the group is different for each momentum sector. 

In general, an irreducible representation restricted to a subgroup
does not remain irreducible. The decomposition of the lowest partial
waves is well known in the literature for all momentum sectors in this
work~\cite{Johnson1982,Mandula1983,Mandula1984,Moore:2005dw,Moore:2006ng,Bernard:2008ax,Gockeler:2012yj}.

The prescription to decompose an eigenstate of the
$l$-th partial wave is often referred to as
\enquote{subduction}. To introduce notation, assume the irrep
$\mathrm{D}^l$ decomposes into a direct sum of different irreps
$\Gamma_i$ each of which appears $n_i$ times such that  
\begin{align}
  \mathrm{D}^l \to \bigoplus_i n_i \Gamma_i \,, \qquad \sum_i \ n_i \cdot \dim(\Gamma_i) = 2l + 1 \,.
  \label{eq:irrep_demomposition}
\end{align}
Let $\Gamma \in \set{\Gamma_i}$, and label the basis vectors of
$\Gamma$ by $\alpha \in \set{1, \dots, \dim(\Gamma)}$. The
decomposition can be completely described by a set of
\enquote{subduction coefficients} denoted by $s$. Given a basis
$\set{\ket{l, m} | -l \leq m \leq l}$, the $\alpha$-th basis vector of
the $n$-th copy of $\Gamma$ is given by 
\begin{equation}
  \ket{\Gamma \alpha l n} = \sum_m s_{lm}^{\Gamma \alpha n} \ket{l m} \,.
  \label{eq:subduced_basis}
\end{equation}
The derivation of subduction coefficients is discussed in appendix
\ref{sec:operator_construction}. Applying the subduction to the matrix
$M$ from \eq{eq:luescher_matrix_lmlm} yields 
\begin{align}
  M_{ln,l'n'}^{\Gamma}
  &= \delta_{\Gamma\Gamma'} \delta_{\alpha\alpha'} \sum_{m m'} {s_{lm}^{\Gamma \alpha n}}^{*} s_{l'm'}^{\Gamma' \alpha' n'} 	
  M_{lm,l'm'} \\
  &= \sum_{m m'} {s_{lm}^{\Gamma \alpha n}}^{*} s_{l'm'}^{\Gamma \alpha n'} 
  (-1)^l \\
  &\quad\times \sum_{j=\abs{l-l'}}^{l+l'} \sum_{s=-j}^{j} \sqrt{2j+1} i^j w_{js} C_{lm,js,l'm'} \,.
  \label{eq:luescher_matrix_lnln}
\end{align}
The Lüscher formula \eq{eq:luescher_formula} remains formally unchanged except
for the space it acts in.
In the following, we will neglect all partial waves apart from the
$p$-wave. In this case $n_i = 1$ for all $i$ and \eq{eq:luescher_formula} simplifies to
\begin{equation}
  \label{eq:lf_red}
  \delta_1 = \arccot{M_{11,11}^{\Gamma}}\,,
\end{equation}
The contributions of higher odd partial waves have been analysed and
found to be negligible~\cite{Dudek:2012xn,Guo:2016zos}. While twisted mass
breaks parity and thus even partial waves may enter, the effect is
suppressed by $\order{a^2}$ and also neglected here.

In \tab{tab:m_lnln_rho} we list the explicit expressions for
$M_{11,11}^{\Gamma}$ used in this work.

\begin{table*}
  \centering
  \begin{tabular}{llc}
    \hline\hline
    $\bm{d}^2$ & $\Gamma$ & $M_{11,11}^{\Gamma}$ \\
    \hline\hline
    $0$	& $\mathrm{T1u}$ & $w_{0,0} - w_{2,0} - 
    \frac{3}{\sqrt{6}} w_{2,-2} - \frac{3}{\sqrt{6}} \cdot w_{2,2}$ \\
    $1$ & $\mathrm{A1}$ & $w_{0,0} + 2 \cdot w_{2,0}$ \\
    $1$ & $\mathrm{E}$ & $w_{0,0} - w_{2,0} + 
    \frac{3i}{\sqrt{6}} \cdot w_{2,-2} - \frac{3i}{\sqrt{6}} \cdot w_{2,2}$ \\
    $2$ & $\mathrm{A1}$	& $w_{0,0} - w_{2,0} +
    \frac{3i}{\sqrt{6}} \cdot w_{2,-2} - \frac{3i}{\sqrt{6}} i \cdot w_{2,2}$ \\
    $2$	& $\mathrm{B1}$ & $w_{0,0} + 2 \cdot w_{2,0}$ \\
    $2$ & $\mathrm{B2}$ & $w_{0,0} - w_{2,0} -
    \frac{3i}{\sqrt{6}} \cdot w_{2,-2} + \frac{3i}{\sqrt{6}} \cdot w_{2,2}$ \\
    $3$ & $\mathrm{A1}$ & $w_{0,0}
    + 2 \cdot \frac{1+i}{\sqrt{6}} \cdot w_{2,-1} - 2 \cdot \frac{1-i}{\sqrt{6}} \cdot w_{2,1}
    + \frac{2i}{\sqrt{6}} \cdot w_{2,-2} - \frac{2i}{\sqrt{6}} \cdot w_{2,2}$ \\
    $3$ & $\mathrm{E}$ & $w_{0,0} -
    \frac{1+i}{\sqrt{6}} \cdot w_{2,-1} + \frac{1-i}{\sqrt{6}} \cdot w_{2,1} -
    \frac{i}{\sqrt{6}} \cdot w_{2,-2} + \frac{i}{\sqrt{6}} \cdot w_{2,2}$ \\
    $4$ & $\mathrm{A1}$ & $w_{0,0} + 2 \cdot w_{2,0}$ \\
    $4$ & $\mathrm{E}$ & $w_{0,0} - w_{2,0} + 
    \frac{3i}{\sqrt{6}} \cdot w_{2,-2} - \frac{3i}{\sqrt{6}} \cdot w_{2,2}$ \\
    \hline\hline
  \end{tabular}
  \caption{Matrix elements for all momentum sectors $\bm{d}^2$ and irreps $\Gamma$ used in
    this work~\cite{Bernard:2008ax}.}
  \label{tab:m_lnln_rho}
\end{table*}

\subsection{Extraction of Energy Levels}

In order to be able to use \eq{eq:lf_red}, we need to extract
interacting energy levels for a given lattice irrep $\Gamma$ as well
as the pion energy $E_\pi(\bm{p})$. The latter is, as usual,
determined from the Euclidean time dependence of two-point functions
\begin{equation}
  \label{eq:Cpi}
  C_\pi(t-t')\ =\ \langle \mathcal{O}_{\Pgpp}(t, \bm{p})^\dagger
  \ \mathcal{O}_{\Pgpp}(t', \bm{p})\rangle
\end{equation}
with operators $\mathcal{O}_{\Pgpp}(t, \bm{p})$ coupling to the
charged pion state with momentum $\bm{p}$, see below. Note that in our 
formulation we have $M_{\pi^+} = M_{\pi^-}$. The spectral
decomposition of $C_\pi$ yields 
\begin{equation}
  C_\pi(t)\ \propto\ \sum_n \left(\eto{-E_n t} + \eto{-E_n\,(T-t)}\right)\,.
\end{equation}
In the limit of large Euclidean times only the ground states survives
and allows one to extract $E_\pi(\bm{p})$ from its exponential decay. 

For irrep $\Gamma$ we define a list of
suitable operators $\mathcal{O}_\Gamma^{i}(t,\mathbf{p})$,
$i=1,\ldots,n$, which project to irrep $\Gamma$ for momentum
$\mathbf{p}$. Because the eigenvalues of operators from the same momentum sector and irrep are degenerate up to statistical fluctuations, we compute the correlator matrix by averaging over all moving frames connected by an allowed lattice rotation and rows of the irrep
\begin{equation}
  \label{eq:gevp}
  \begin{split}
    \mathcal{C}_{\Gamma,
      \bm{d}^2}(t-t')\ &=\ \frac{1}{\abs{\set{\bm{d}}}} \sum_{\bm{p}
      \ \in \ \set{\bm{d}}} \frac{1}{\dim(\Gamma)} \\
    &\quad \times \sum_{\alpha = 1}^{\dim(\Gamma)} \langle\ \vec{\mathcal{O}}_{\Gamma}^{\alpha}(t,
    \mathbf{p})^\dagger
    \cdot  \vec{\mathcal{O}}_{\Gamma}^{\alpha}(t',
    \mathbf{p})\ \rangle\,,\\
  \end{split}
\end{equation}
where we defined $\vec{\mathcal{O}}_{\Gamma}^{\alpha} = (\mathcal{O}_{\Gamma 1}^{\alpha},
\ldots, \mathcal{O}_{\Gamma n}^{\alpha})^t$. The correlator matrix
$\mathcal{C}_{\Gamma \mathbf{d}^2}(t)$ is then analysed using the standard variational
method~\cite{Michael:1982gb,Luscher:1990ck} yielding eigenvalues
$\lambda_i(t, t_0)$ which, at large enough $t$-values, decay like
\begin{equation}
  \label{eq:lambda}
  \lambda_i(t, t_0)\ \propto \exp(-W_i(t-t_0)) + \exp(-W_i(T - t+t_0))\,,
\end{equation}
where we neglect thermal pollutions for the moment, see
section~\ref{sec:thermalState}. 
Here, $T$ is the time extent of the lattice and $W_i$ the $i$th energy
level to be extracted. $t_0$ represents the reference time at which the
generalised eigenvalue problem (GEVP) is seeded. The correction
to \eq{eq:lambda} due to excited states
reads at fixed $t_0$-value~\cite{Luscher:1990ck}
\begin{equation}
  \label{eq:excst}
  \varepsilon_i(t,t_0)=O(e^{-\Delta W_it})\,.
\end{equation}
Here, $\Delta W_i$ is the energy difference of $W_i$ to the first
state not resolved by the correlation matrix. For a detailed
discussion see Ref.~\cite{Blossier:2009kd}. 

\subsection{Operator Construction}
\label{sec:operators}

We start with interpolating operators for pions $\pi^\pm$ with
definite isospin $\ket{1, \pm1}_{I}$:
\begin{equation}
  \begin{split}
    \mathcal{O}_{\Pgpp}(x) &= \bar{d}(x)_\alpha^c \Gamma^{\pi}_{\alpha\beta} u(x)_\beta^c \,,\\
    \mathcal{O}_{\Pgpm}(x) &= \bar{u}(x)_\alpha^c
    \Gamma^{\pi}_{\alpha\beta} d(x)_\beta^c \,,
  \end{split}
  \label{eq:operator_Pgp}
\end{equation} 
where $u$ and $d$ denote Dirac spinors for an up and down quark,
respectively. $\alpha$, $\beta$ denote spin and $c$ colour indices, and
$\Gamma^{\pi}=i\gamma_5$.

For the $\rho$-meson, we have to construct operators projected to
$I=1$. A single $\rho^0$ can be interpolated by the canonical anti symmetric
combination of quarks with isospin $\ket{1, 0}_{I}$: 
\begin{align}
  \mathcal{O}_{\Pgr}(x) = \frac{1}{\sqrt{2}} (\bar{u}(x)_\alpha^c \Gamma^{\Pgr}_{\alpha\beta} u(x)_\beta^c - 
  \bar{d}(x)_\alpha^c \Gamma^{\Pgr}_{\alpha\beta} d(x)_\beta^c)\,.
  \label{eq:RhoBilinear}
\end{align}
$\Gamma^{\Pgr}$ must ensure that $\mathcal{O}_{\Pgr}$ transforms like
$J^{PC} = 1^{--}$, i.e. $\Gamma^{\Pgr}\in\{i\gamma_i,
\gamma_0\gamma_i\}$. From the operators for charged pions
\eq{eq:operator_Pgp} one can construct two pion operators with $I=1$
as follows
\begin{equation}
  \begin{split}
  \mathcal{O}_{\Pgp\Pgp}(t, \bm{x}_1, \bm{x}_2)
  &= \frac{1}{\sqrt{2}} \left( \mathcal{O}_{\Pgpp}(t, \bm{x}_1) \mathcal{O}_{\Pgpm}(t, \bm{x}_2)\right.\\  
  &\quad- \left.\mathcal{O}_{\Pgpm}(t, \bm{x}_1) \mathcal{O}_{\Pgpp}(t, \bm{x}_2)
  \right)\,.\\
  \end{split}
  \label{eq:operator_PgpPgp}
\end{equation}
The projection of a given single particle operator $\mathcal{O}(t,
\bm{x})$ to momentum $\bm{p}$ is performed via
\begin{equation}
  \mathcal{O}(t,
  \bm{p})\ =\ \sum_{\bm{x}}\ \mathcal{O}(t, \bm{x})\,
  e^{i\bm{x}\bm{p}} 
\end{equation}
and likewise for two particle operators $\mathcal{O}(t, \bm{x}_1,
\bm{x}_2)$ to momenta $\bm{p}_1$, $\bm{p}_2$, respectively,
yielding $\mathcal{O}_{\pi\pi}(t, \bm{p}_1+\bm{p}_2)$.

The projection to a given lattice irrep $\Gamma$ and basis vector
$\alpha$ is performed via the so-called subduction procedure described
in appendix~\ref{sec:operator_construction}. 

\subsection{Thermal State Pollution}
\label{sec:thermalState}

Apart from excited state contaminations \eq{eq:excst} there are
additional so-called thermal state pollutions, which are relevant with
finite time extent $T$, periodic boundary conditions and in the
presence of multi-particle states.

For the case of pion-pion systems with momenta $\bm{p}_{1,2}$, the
leading thermal pollution to a matrix element of the correlator matrix
$\mathcal{C}^{\Gamma\alpha}$ reads
\begin{equation}
  \label{eq:epst}
  \begin{split}
    \varepsilon_t(t, \bm{p}_1, \bm{p}_2)\ &\propto\
    \eto{-E_{\Pgp}(\bm{p}_1) T} \eto{-(E_{\Pgp}(\bm{p}_2) -E_{\Pgp}(\bm{p}_1)) t}\\
    &\quad + \eto{-E_{\Pgp}(\bm{p}_2) T} \eto{-(E_{\Pgp}(\bm{p}_1)
      -E_{\Pgp}(\bm{p}_2)) t}\,.
  \end{split}
\end{equation}
For $\bm{p}_1 = \bm{p}_2$ this is a constant contribution and the time
dependence drops out. The thermal pollution $\varepsilon_t$ vanishes
for $T\to\infty$. However, at finite $T$ it can become relevant
for $t\to T/2$. There are, of course, further pollution terms which
are exponentially suppressed compared to the one quoted above.
Let us now assume $E_{\Pgp}(\bm{p}_2)>E_{\Pgp}(\bm{p}_1)$ and 
concentrate on the corresponding, exponentially decreasing term in
\eq{eq:epst}. This is sufficient because the signal to noise ratio
in the relevant correlator matrices is decreasing exponentially with
Euclidean time. Therefore, we will have to extract the signal at
relatively small $t$-values where the second, exponentially increasing
term in $\varepsilon_t$ is not yet relevant.

We can deal with this pollution term by applying the so-called
weighting and shifting procedure~\cite{Dudek:2012gj}. It amounts to the following
transformation of $\mathcal{C}$:
\begin{equation}
  \label{eq:Ctilde}
  \tilde{\mathcal{C}}(t)\ =\ \eto{-\Delta E\, t}\left(\mathcal{C}(t)\eto{\Delta E\,
    t} - \mathcal{C}(t+1)\eto{\Delta E\,(t+1)}\right) \,,
\end{equation}
with $\Delta E = E_{\Pgp}(\bm{p}_2)-E_{\Pgp}(\bm{p}_1)$. It is easy to
see that this transformation leaves the leading, physical exponential
dependence unchanged, while the thermal pollution is removed. As an
input for the transformation \eq{eq:Ctilde} we use $E_\pi(\bm{p})$
determined from single charged pion two point functions at zero
momentum combined with the continuum dispersion relation \eq{eq:disp}. 

We remark here that we have investigated thermal pollutions in some
detail in Ref.~\cite{Helmes:2018nug}. However, the corresponding
findings are not applicable here, because the signal does not extend
to large enough $t$-values.

\subsection{Phase Shift Curves}

Once the energy levels have been determined for all the irreps
mentioned above, the phase shift $\delta_1$ is to be determined from
\eq{eq:lf_red}. This requires the evaluation of the Lüscher zeta
function $\mathcal{Z}_{lm}(1, q^2)$ in $w_{lm}$. $\mathcal{Z}$ has
poles at $q^2$-values corresponding to the free, non-interacting two
particle energies. The larger the spatial extent $L$ of the lattice,
the closer are the interacting energy levels to these poles.

This structure makes the error estimate for $\delta_1$ difficult in
cases where the statistical uncertainty of the interacting energy
levels is not small enough: when an energy level is compatible with a
pole of the $\mathcal{Z}$-function within errors, a proper estimate 
of the uncertainty of $\delta_1$ becomes impossible. However, also
when this is not the case, such a situation can still be and actually
is triggered in some cases during a
bootstrap analysis. Since bootstrap replicates are sampled uniformly
random with replacement, it is not unlikely to hit a pole of the
$\mathcal{Z}$-function, even if the pole is two or three sigma away
from the actual energy level.

To circumvent this problem, we use instead of the bootstrap the
jack-knife procedure, which can be understood as a linear
approximation to the bootstrap. The standard-deviation over jack-knife
replicates is per construction a factor of $\sqrt{N-1}$ smaller than
the one over bootstrap replicates, where $N$ is the sample size. 

It is clear that using the jack-knife procedure introduces
additional uncertainties due to the linearisation, in particular in
the vicinity of a singularity of the $\mathcal{Z}$-function. We have
compared the jack-knife and bootstrap procedure for all cases, where
bootstrap did not show the aforementioned problem. For all these cases
we found excellent agreement for the error estimate between the two
methods. Thus, we conclude that the systematic error introduced by
jack-knife is likely not significant, even though we cannot make this
statement definite. 

With this procedure we then determine $\delta_1$ as a function of
$E_\mathrm{CM}$ using equation \eq{eq:lf_red}. The next step is to
determine the $\rho$-meson mass $M_\rho$ and width $\Gamma_\rho$
from these phase shift points. For this purpose we use a relativistic
Breit-Wigner functional form 
\begin{equation}
  \label{eq:tan_delta}
  \begin{split}
    \tan\delta_1 &= \frac{g_{\rho\pi\pi}^2}{6\pi}
    \frac{p^3(E_\mathrm{CM})}{E_\mathrm{CM}\cdot(M_\rho^2 -
      E_\mathrm{CM}^2)}\,,\\
    p(E_\mathrm{CM}) &=
    \sqrt{E_\mathrm{CM}^2/4 - M_\pi^2}\,,
  \end{split}
\end{equation}
which we fit to our data. Here, $g_{\rho\pi\pi}$ is the
$\rho$ to $\pi\pi$ coupling constant. The width is related to
$g_{\rho\pi\pi}$ through $M_\rho$ via
\begin{equation}
  \Gamma_{\Pgr} = \frac{2}{3} \frac{g_{\Pgr\Pgp\Pgp}^2}{4\pi} \frac{p^3(M_{\Pgr})}{M_{\Pgr}^2} \,.
  \label{eq:rho_width}
\end{equation}
\eq{eq:tan_delta} allows to extract the mass and width from the phase
shift data at a given pion mass. We remark that \eq{eq:tan_delta}
contains several approximations. The resonance must be isolated and
narrow. Additionally $\tan \delta_1$ has a pole at $E_\mathrm{CM} =
M_{\Pgr}$ which was rewritten as a rational function where the denominator is a first-order polynomial in $k^2$. 
For $M_{\Pgr} = 775 \ \mathrm{MeV}$ the predicted width is 
$\Gamma_{\Pgr} \simeq 130 \ \mathrm{MeV}$~\cite{Brown:1968zza}. Additional modifications such 
as barrier terms, have been observed to slightly improve fit quality, but 
had no significant effect on the final results. \cite{Alexandrou:2017mpi,Bali:2015gji,Dudek:2012xn} 

Since $N_\textrm{conf}$ is different on all our ensembles, the jack-knife procedure 
is not easily applied in such a chain
of analyses and we take the jack-knife errors as an input to a parametric
bootstrap procedure. Here we generate the parametric bootstrap
replicates such as to have the same correlation between
$E_\mathrm{CM}$, $M_\pi$ and $\delta_1$ as the jack-knife
replicates. Then we fit \eq{eq:tan_delta} to our data for $E_\mathrm{CM}$, $\delta_1$
and $M_\pi$ with two free parameters $g_{\rho\pi\pi}$ and $M_\rho$.

\subsection{Pion Mass Dependence}

In Ref.~\cite{Bruns:2004tj} the pion mass dependence of the
$\rho$-meson mass has been computed using effective field theory with
infrared regularisation. Up to $\mathcal{O}(M_\pi^3)$ plus the
non-analytic term of order $M_\pi^4$, the dependence reads
\begin{equation}
  \label{eq:brunsmeissner}
  M_\rho(M_\pi^2)\ =\ M_\rho^0 + c_1 M_\pi^2 + c_2 M_\pi^3 + c_3
  M_\pi^4\ln\left(\frac{M_\pi^2}{M_\rho^2}\right) + \mathcal{O}(M_\pi^4)\,.
\end{equation}
To this order the formula contains four unknown parameters, the $\rho$
mass in the chiral limit $M_\rho^0$ and the parameters $c_1, c_2$ and
$c_3$. Using this mass dependence of $M_\rho$ and the KSFR
relation~\cite{Kawarabayashi:1966kd,Riazuddin:1966sw}, 
we can try to relate $g_{\rho\pi\pi}$ to $M_\pi$ up to order 
$M_\pi^3$ using \eq{eq:brunsmeissner} and the SU$(2)$ chiral perturbation
theory formula for $f_\pi$~\cite{Gasser:1983yg}
\begin{equation}
  \label{eq:grpipi}
  \begin{split}
    &g_{\rho\pi\pi}(M_\pi^2)\ \approx\ \frac{M_\rho}{f_\pi}\\
    &\approx
  \frac{1}{f_0}\left[M_\rho^0 + M_\pi^2\left(c_1 + \frac{2}{16\pi^2
      f_0^2}(\log\xi_\ell -\bar\ell_4 - \ell_\pi)\right)\right.\\
    &\qquad\quad\left. +c_2 M_\pi^3
    \right]+ \mathcal{O}(M_\pi^4)\,. \\
  \end{split}
\end{equation}
Here, $f_\pi$ is the pion decay constant, $f_0$ its value in the
chiral limit and the parameters $M_\rho^0$ and $c_i$ are the ones from
\eq{eq:brunsmeissner}. Note that we follow the convention with
$f_\pi\approx 130\ \mathrm{MeV}$~\cite{Tanabashi:2018oca}. In addition
we have used the definitions 
\[
\ell_\pi\ =\ \log\left(\frac{\overline{M}_{\pi^+}}{4\pi
  f_0}\right)^2\,,\quad \xi_\ell\ =\ \frac{M_\pi^2}{16\pi^2 f_0^2}
\]
and the usual low energy constant $\bar\ell_4$.
Values for $f_0$ and $\bar\ell_4$ have
been computed on the ensembles used here in
Ref.~\cite{Carrasco:2014cwa}
\[
f_0\ =\ 121.1(2)\ \mathrm{MeV}\,,\qquad \bar\ell_4\ =\ 4.7(1)\,.
\]
We remark that the KSFR
relation~\cite{Kawarabayashi:1966kd,Riazuddin:1966sw}
$g_{\rho\pi\pi} \approx M_\rho/f_\pi$ 
is fulfilled in nature to very good approximation. However, it is not
clear at all whether it can be extended beyond leading order in the
pion mass.

In Ref.~\cite{Djukanovic:2009zn,Djukanovic:2010id}, the pion mass dependence
of the \Pgr-meson mass and width has been calculated with the
complex mass renormalisation scheme from an effective field theory
with explicit contributions corresponding to the \Pgo-meson. It is
based on the assumption of vector meson dominance and, thus, model
dependent; see also Ref.~\cite{Ecker:1989yg} for details on the
model. However, its advantage is that mass and width can be
extrapolated in a combined fit. The
squared pole position of the $\Pgr$ resonance, $Z = \left( M_{\Pgr} -
\i/2 \ \Gamma_{\Pgr} \right)^2$ has the following pion mass dependence
\begin{align}
Z = Z_\chi 
+ c_\chi M_{\Pgp}^2 
- \frac{g^2_{\Pgo \Pgr \Pgp}}{24 \pi} Z_\chi^{\nicefrac12} M_{\Pgp}^3
+ \order{M_{\Pgp}^4}\,,
\end{align}
where $Z_{\chi}$ is the pole position in the chiral limit and
$c_{\chi}$, $g_{\Pgo\Pgr\Pgp}$ are coupling constants. Higher order
corrections in $M_{\Pgp}$ are known in principle, which also include
logarithmic terms. The non-analytic structure in $M_\rho$ is identical
to the one of \eq{eq:brunsmeissner}.

In order to apply this formula to our lattice data, we re-express it
in units of the Sommer parameter $r_0$
\begin{equation}
  \begin{split}
    \sommer^2 Z &= \sommer^2 Z_\chi
    + C_\chi (\sommer M_{\Pgp})^2\\
    &\quad - \frac{g^2_{\Pgo \Pgr \Pgp}}{24 \pi \sommer^2} 
  (\sommer^2 Z_\chi)^{\nicefrac12} (\sommer M_{\Pgp})^3
    + \frac{p_{a^2}}{\sommer^2} a^2\\
  \end{split}
  \label{eq:feng_eq19_sommer}
\end{equation}
and add an $a^2$ term, which represents the leading lattice artefacts
for the twisted mass formulation at maximal twist.
$p_{a^2}$ is an unknown complex parameter.

\begin{figure}[th]
  \centering
  \subfigure{\includegraphics[width=0.47\textwidth]{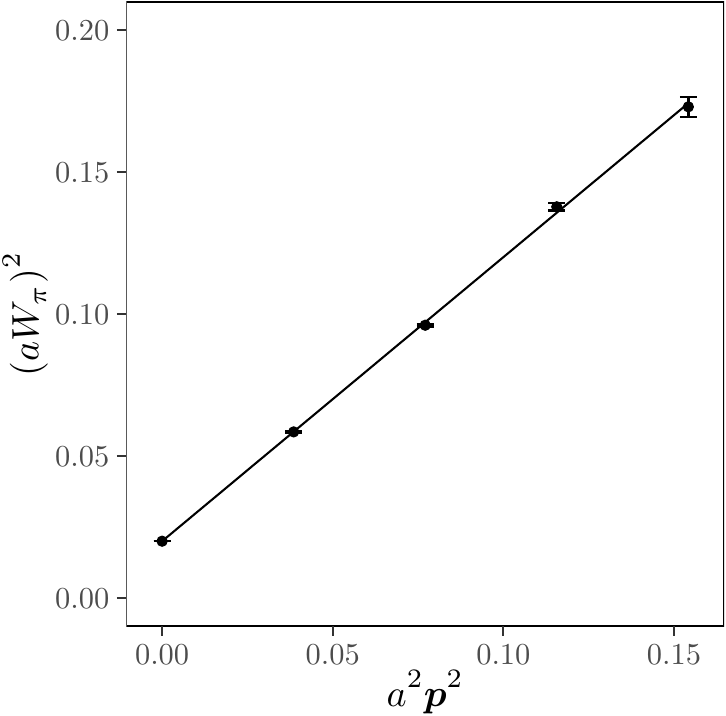}}\quad
  \subfigure{\includegraphics[width=0.475\textwidth]{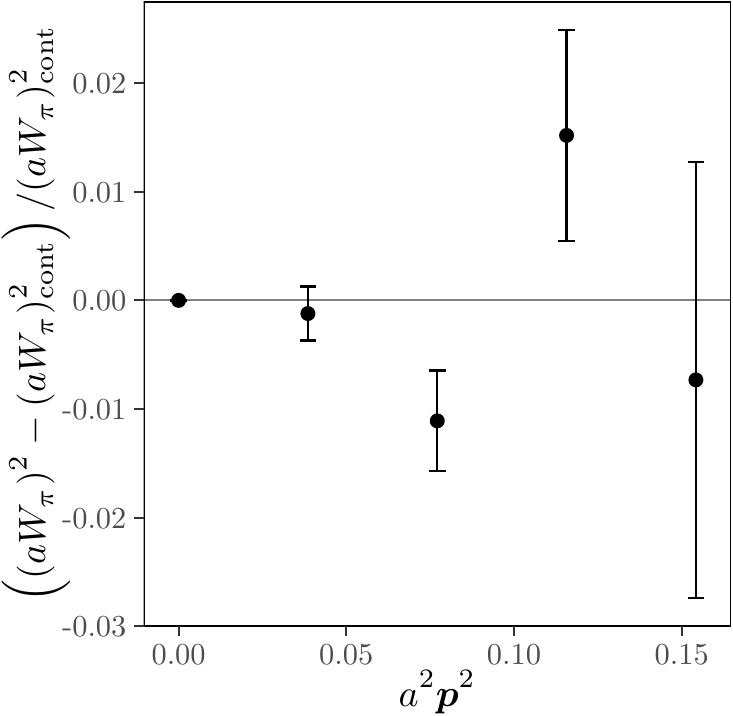}}
  \caption{Dispersion relation of the pion for ensemble A40.32. Left:
    $W_\pi(\bm{p})^2$ as a function of $\bm{p}^2$, both in lattice
    units. The solid line represents the continuum dispersion
    relation. Right: relative difference of measured $W_\pi(\bm{p})^2$
    and the corresponding prediction of the continuum dispersion relation.} 
  \label{fig:dispersion}
\end{figure}

\section{Results}
\label{sec:results}

\subsection{Pion Dispersion Relation}

In order to extract the energy shift, we need the pion energy
not only at rest but also in moving frames. As mentioned before,
in order to reduce statistical uncertainties we are going to use the
relativistic continuum dispersion relation 
\begin{equation}
  \label{eq:dispersion}
  W_\pi^2(\bm{p}) = M_{\Pgp}^2 + \bm{p}^2
\end{equation}
to compute $W_\pi(\bm{p})$ from the zero momentum pion mass value. As
a check for the validity of this approach we have also computed
$W_\pi(\bm{p})$ from two-point correlation functions with momentum.

In \fig{fig:dispersion} we compare the measured $W_\pi^2(\bm{p})$ with the
prediction of \eq{eq:dispersion} with $M_\pi^2$ at zero momentum as
input exemplarily for the A40.32 ensemble. Good agreement within
errors is observed up to $\bm{d}^2=4$. This makes us confident that
using the dispersion relation is safe.

\subsection{Energy Levels}

One of the major uncertainties in our extraction of energy levels of
multi-particle correlation functions is caused by thermal
pollutions. For the case of two pions with maximal isospin the onset
of thermal pollutions in Euclidean time in the correlators is clearly
visible. However, due to the exponential deterioration of the
signal-to-noise ratio, this is not the case for the correlation
functions investigated here. This manifests itself also in the fact
that there is no clear difference visible between principal
correlators $\lambda(t, t_0)$ derived from $\mathcal{C}(t, t_0)$ or their
weighted and shifted counterparts $\tilde{\lambda}(t,t_0)$ derived
from $\tilde{\mathcal{C}}(t, t_0)$. Therefore, we perform the full
analysis with and without weighting and shifting and include the
difference as a systematic uncertainty in our error budget.

\begin{figure}[th]
  \centering
  \subfigure{\includegraphics[width=.48\textwidth]{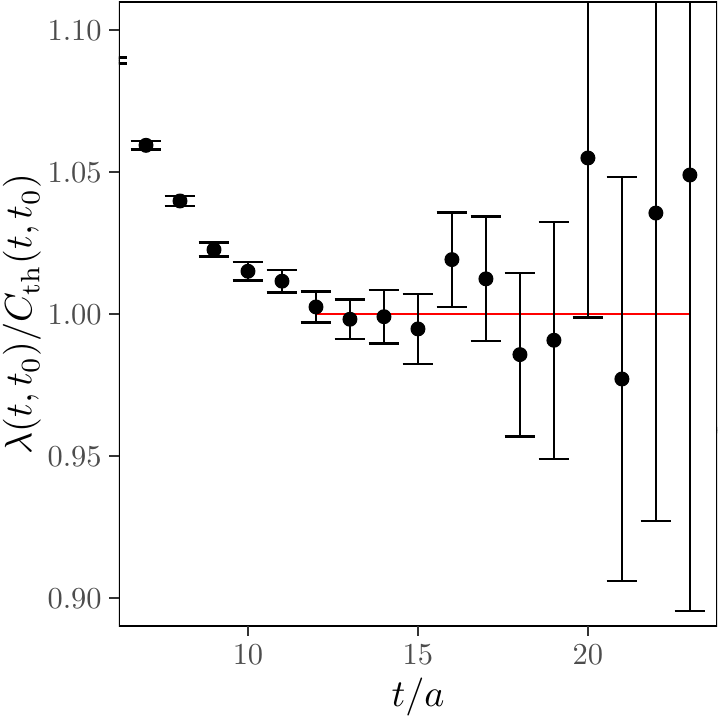}}
  \subfigure{\includegraphics[width=.48\textwidth]{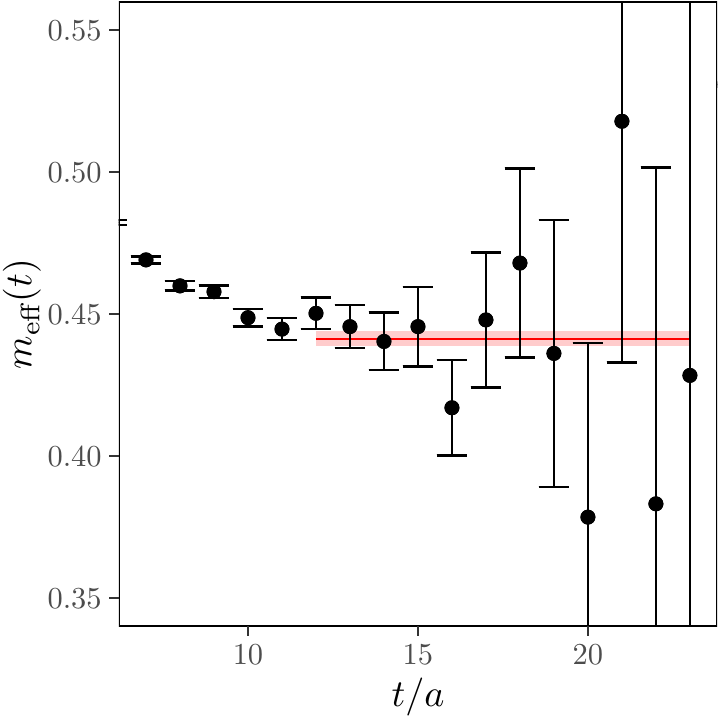}}
  \caption{In the left panel we show $\lambda(t, t_0)/C_\mathrm{th}(t, t_0)$ 
    as a function of $t/a$ for the ground state energy level in irrep $\mathrm{E}$. 
    The reference time for the GEVP was set to $t_0/a = 3$ and the ensemble is A40.32. 
    The horizontal line indicates the fit range. 
    In the right panel we show the effective mass as a function of $t/a$ and the fitted 
    energy value with error band for reference.}
  \label{fig:lambda1}
\end{figure}

\begin{figure}
  \centering
  \subfigure{\includegraphics[width=.48\textwidth]{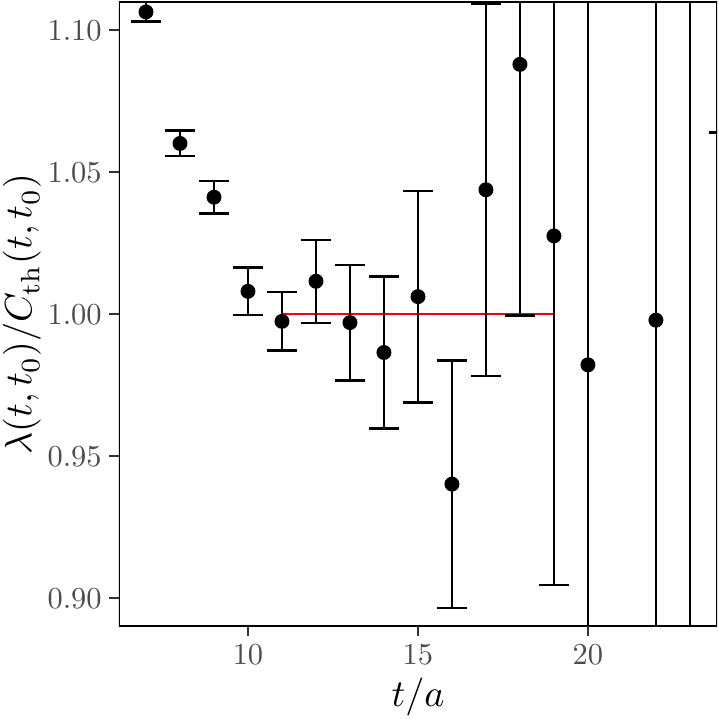}}
  \subfigure{\includegraphics[width=.48\textwidth]{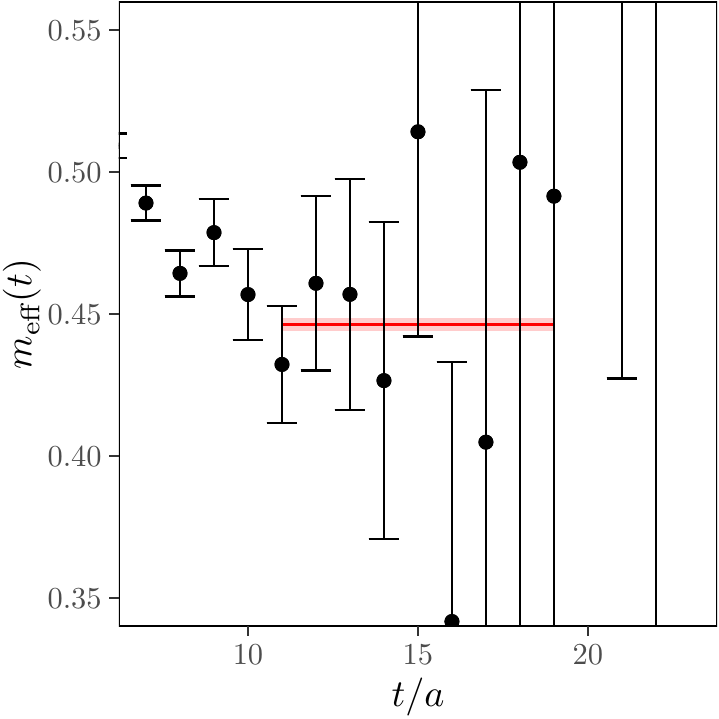}}
  \caption{the same as \fig{fig:lambda1}, but for weighted and shifted
    $\tilde{\lambda}$.} 
  \label{fig:lambda2}
\end{figure}

The other major uncertainty in extracting energy levels from
lattice correlation functions stems from the choice of fit
range. There have been approaches to make this choice more objective by
performing a weighted average over many fit ranges, which works well
for the case of single pions 
or two pions with maximal isospin. In contrast, for the case in
question here, the $\rho$ channel, the weighted average turns out not to
be useful.

Therefore, our procedure is the following: we perform the fitting to
the principal correlator $\lambda(t, t_0)$ (and $\tilde{\lambda}$) by
surveying multiple fit ranges $[t_\text{min}, t_\text{max}]$ 
and selecting a representative one. We enforce a plateau length of at
least four points, which must be compatible within errors and have
relative errors below 50\%. Additionally we require no significant 
dependence on $t_\text{max}$ as this would be a consequence of residual 
thermal pollution. The dependence on $t_\text{min}$ is very pronounced
when $t_\text{min}$ is in a region, where excited states are still
relevant. We increase $t_\text{min}$ until this dependence vanishes.
A $p$-value above $0.05$ was preferred
to ascertain that the data in the chosen range are described by our
fit. In the rare cases where multiple fit ranges gave competing 
and equally likely results, we chose an intermediate range. The influence
of varying $t_0$ from $1$ to the onset of the plateau was checked and
found to be negligible. Therefore, we chose  
$t_0 = 3$ on the coarser two and $t_0 = 4$ on the finest lattice
spacing, corresponding to  
approximately $0.25 \ \textrm{fm}$ in physical units. Finally, all other
qualities being equal, we preferred larger $t_\text{max}$.

\begin{figure*}
  \centering
  \includegraphics[width=.85\textwidth]{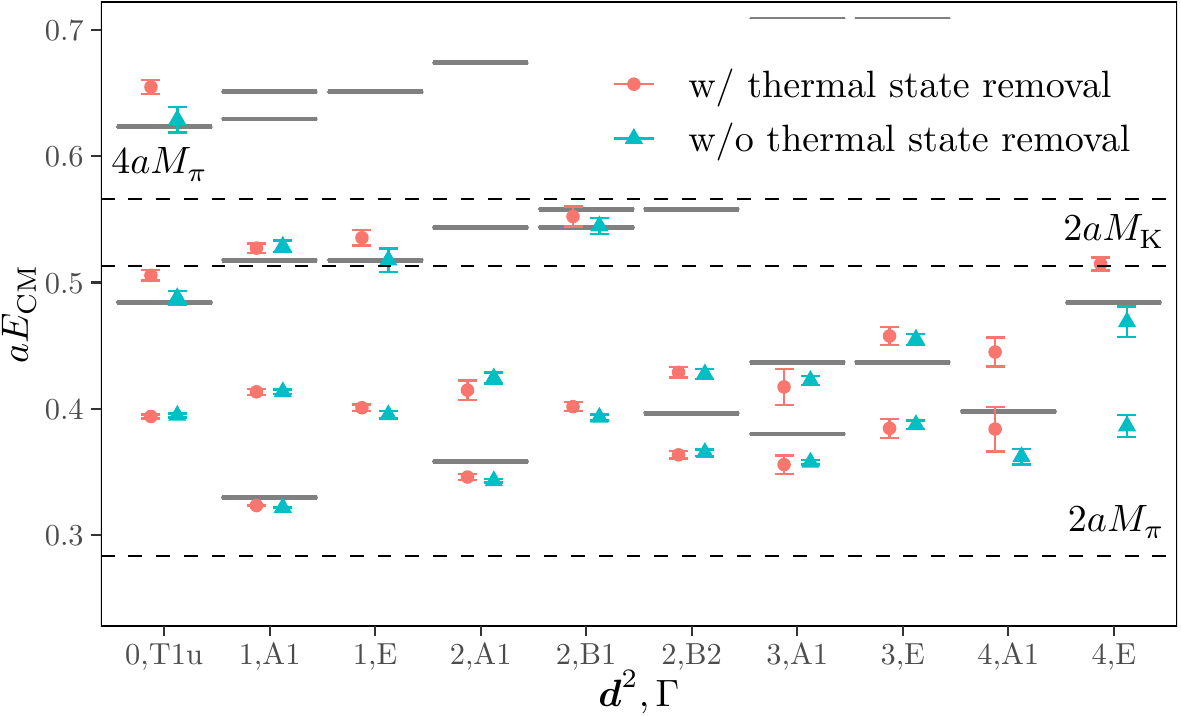}  
  \caption{Example of all energy levels in lattice units for ensemble
    A40.32 for irrep $\Gamma$ and $\pcm$ labeled by $\bm{d}^2$. The two
    kaon, two pion and four pion thresholds are indicated by the dashed 
    horizontal lines. The shorter solid lines indicate the
    non-interacting energy levels in each irrep.
    The two colours and symbols distinguish the estimate of $E_\mathrm{CM}$
    with and without thermal state removal.}
  \label{fig:spectrum}
\end{figure*}

In \fig{fig:lambda1} we show an example for the fit range chosen for ensemble 
A40.32 where $d^2 = 1$ and irrep $\Gamma = \textrm{E}$ without weighting and shifting. 
In the left panel, we show the ratio of principal correlator $\lambda(t, t_0)$ 
and the single exponential fit model 
$\mathcal{C}_\mathrm{th}(t, t_0) = \exp(- W (t-t_0))$. Compared to the effective 
mass, the ratio is more robust numerically. By definition the central value is $1$. In
the right panel we show for illustration the result of the correlator fit as a red band 
along with the effective mass
\begin{equation*}
  m_\text{eff}(t) = \log \frac{\mathcal{C}(t)}{\mathcal{C}(t+1)} \,.
\end{equation*}
As mentioned above, the effects of thermal states 
are not visible here. The energy level was determined as $aW = 0.4412(26)$. 

In \fig{fig:lambda2} we show the same plots but this time with
weighting and shifting. The size of error bars is increased compared to
without weighting and shifting, which can be explained by the reduced
correlation of neighbouring time slices. For very large $t$, points are
not depicted because they were compatible with zero. For this reason, 
$t_\text{max}$ was chosen smaller compared to before. 
The fit model was modified as described in \eq{eq:Ctilde} 
and the calculation of the effective mass in the right panel was changed accordingly.
The fit result increased by roughly one standard deviation to $aW = 0.4463(23)$. 
Whether this results from the independent choice of a fit range or due
to not visible but barely significant thermal states remains hard to
decide. By including this difference as a systematic error we are
confident that we keep control of both major sources of systematic
uncertainties. 

In \fig{fig:spectrum} we show all energy levels $aE_\mathrm{CM}$ for
all irreps $\Gamma$ and boosts $\bm{d}^2$ exemplary for ensemble
A40.32. The 
red circles are with weighting and shifting, the blue triangles without. The
two kaon upper and two pion lower thresholds are indicated by the
dashed horizontal lines. For all $\bm{d}^2$-value and irrep combinations,
apart from two, we have two energy levels below the two kaon inelastic
threshold. 

Comparing energy levels with and without thermal state removal, we
observe good agreement. Statistical uncertainties are in general
larger with weighting and shifting. 

\subsection{Phase Shift Determination}

\begin{figure*}[t]
  \centering
  \includegraphics[width=.85\textwidth]{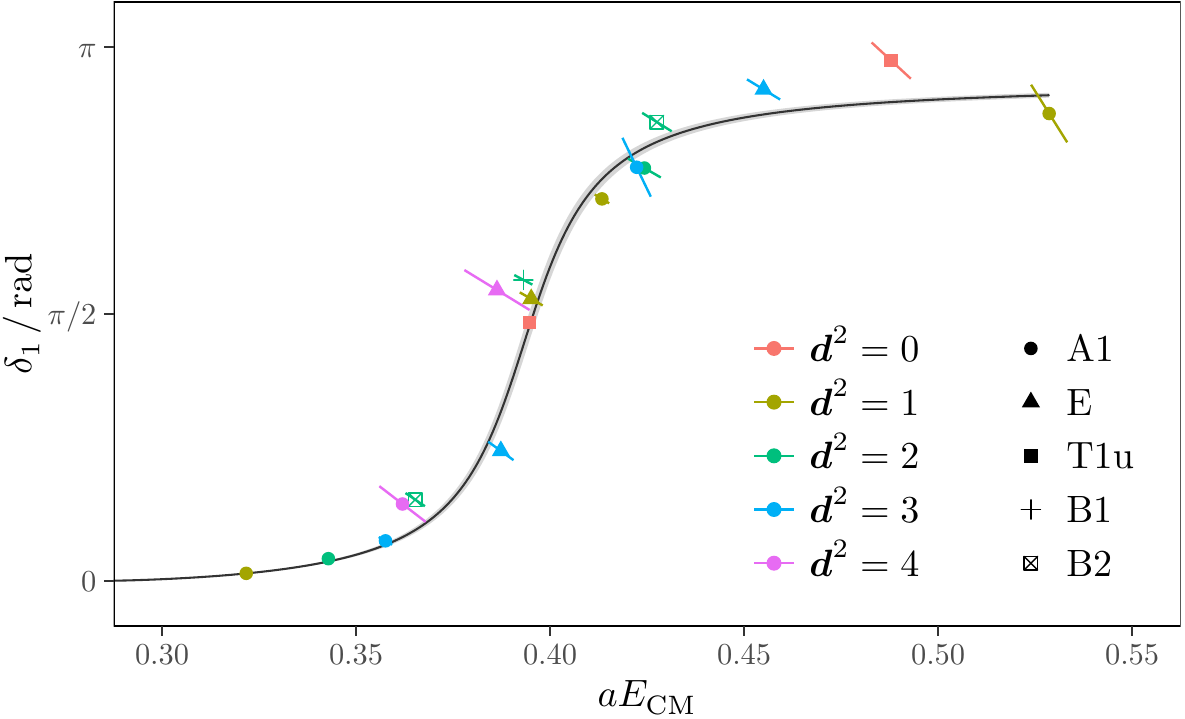}
  \caption{Phase shift $\delta_1$ as a function of $E_\mathrm{CM}$ in
    lattice units for ensemble A40.32. The solid line with error
    band represents the fit result of \eq{eq:tan_delta} to all the
    data w/ thermal state removal. Colours encode the different
    $\bm{d}^2$-values, while symbols distinguish the
    irreps.}
  \label{fig:delta1}
\end{figure*}

In \fig{fig:delta1} we show the phase shift $\delta_1$ as a function
of the centre-of-mass energy $aE_\mathrm{CM}$ for ensemble A40.32. The
two-parameter fit of \eq{eq:tan_delta} to our data is 
shown as a solid line with error band. Colours and symbols encode
$\bm{d}^2$-values and irreps $\Gamma$, respectively. Error bars for
the data points are slanted: $x$- and $y$-errors are added
vectorially, i.e. the length of the slanted error bars is the sum of
$x$- and $y$-error added in quadrature. Positive or negative slope of
the slanted error bar indicates positive or negative correlation
between $x$- and $y$-data. From \fig{fig:delta1} one can, hence,
deduce that $\delta_1$ is negatively correlated with $aE_\mathrm{CM}$.
Note that for determining $\delta_1$ also $aM_\pi$ is needed. Here we
use the finite volume estimate as argued in Ref.~\cite{Romero-Lopez:2018rcb}.

One also reads off from \fig{fig:delta1} that our fit describes the
data particularly well in the region where $\delta_1$ passes through
$\pi/2$. Larger deviations can be observed for larger values of
$\delta_1$, which significantly increase the $\chi^2$-values.

We have performed a list of variations of the fit to the phase shift
data: a) the fits are being performed with and without (w and w/o)
thermal state removal; b) we have performed fits by removing all
points with $\bm{d}^2 > k$ with $k=3, 2, 1$. While b) merely
influences the statistical uncertainty, a) leads to up to $4$ standard
deviations differences in the fit parameters, in particular in $M_\rho$. 
However, it is not clear 
whether approaches w/o or w/ thermal state removal are systematically
cleaner: in the former case we might be plagued with thermal state
pollutions, while in the latter case the fit range might be 
chosen incorrectly due to noise.

Therefore, we decided to use the weighted mean over results w/o and w/
thermal state removal. In addition we include the difference
$\Delta Q_Y$ between the weighted mean and w/o and w/ thermal state
removal into the error by rescaling the bootstrap distribution with a
factor~\cite{Helmes:2018nug} 
\begin{equation}
  s\ =\ \sqrt{\frac{(\Delta x)^2 + \sum_{Y} (\Delta Q_{Y})^2}{(\Delta x)^2}}\,.
\end{equation}
Here, $\Delta x$ is the statistical uncertainty of the weighted mean
and $Y\in\{\mathrm{w/o}, \mathrm{w/}\}$. 

All results for $M_\rho$ and $g_{\rho\pi\pi}$ determined by this
procedure w/ and w/o thermal state removal are compiled in
\tab{tab:rho_mass}. The width $\Gamma_\rho$ computed via
\eq{eq:rho_width} is tabulated in \tab{tab:width}. In 
the latter table we also give the reduced $\chi^2$-values of the
Breit-Wigner fits and the
values for the (charged) pion mass in lattice units $aM_\pi$.

\begin{table*}[ht]
  \centering
  \begin{tabular*}{.8\textwidth}{@{\extracolsep{\fill}}lrrrrrr}
    \hline\hline
    Ensemble & $aM_\rho^\mathrm{w/o}$ & $aM_\rho^\mathrm{w/}$ &
    $aM_\rho^\mathrm{av}$ &
    $g_{\rho\pi\pi}^\mathrm{w/o}$ &
    $g_{\rho\pi\pi}^\mathrm{w/}$ & $g_{\rho\pi\pi}^\mathrm{av}$ 
    \\ 
    \hline\hline
    A30.32  & 0.3906(11) & 0.3968(15) & 0.3929(32) & 6.0(2) & 5.8(2) & 6.0(2) \\
    A40.24  & 0.4010(15) & 0.4084(14) & 0.4051(38) & 5.7(1) & 4.9(2) & 5.4(4) \\
    A40.32  & 0.3957(12) & 0.3971(13) & 0.3964(11) & 5.7(1) & 5.5(2) & 5.6(1) \\ 
    A60.24  & 0.4134(12) & 0.4170(12) & 0.4153(20) & 5.4(1) & 5.4(1) & 5.4(1) \\ 
    A80.24  & 0.4265(11) & 0.4314(14) & 0.4282(26) & 5.3(1) & 5.0(3) & 5.2(2) \\
    A100.24 & 0.4512(11) & 0.4521(12) & 0.4516(09) & 4.7(2) & 5.0(2) & 4.9(2) \\ 
    \hline
    B25.32  & 0.3527(30) & 0.3608(40) & 0.3556(47) & 6.3(3) & 5.9(6) & 6.2(4) \\
    B35.32  & 0.3554(17) & 0.3582(17) & 0.3568(18) & 6.3(2) & 5.4(3) & 6.0(5) \\
    B35.48  & 0.3617(15) & 0.3609(26) & 0.3615(13) & 5.8(2) & 6.6(5) & 6.0(4) \\ 
    B55.32  & 0.3709(09) & 0.3739(09) & 0.3722(16) & 5.6(1) & 6.1(1) & 5.8(3) \\
    \hline
    D15.48  & 0.2751(35) &          - & 0.2751(35) & 6.5(7) &      - & 6.5(7) \\ 
    D30.48  & 0.2747(16) & 0.2926(22) & 0.2811(91) & 5.3(4) & 5.1(5) & 5.2(3) \\ 
    D45.32  & 0.2866(09) & 0.2948(14) & 0.2890(42) & 5.8(2) & 4.6(5) & 5.6(6) \\ 
    \hline\hline
  \end{tabular*}
  \caption{$\rho$ mass $aM_\rho$ and coupling $g_{\rho\pi\pi}$ for all
    ensembles w/ and w/o thermal state 
    removal and the weighted average including the systematic
    uncertainty as explained in the text.}
  \label{tab:rho_mass}
\end{table*}
 \begin{table*}[ht]
  \centering
  \begin{tabular*}{1.\textwidth}{@{\extracolsep{\fill}}lrrrrrrr}
    \hline\hline
    Ensemble & $aM_\pi$ & $K_{M_\pi}$ &
    $a\Gamma_\rho^\mathrm{w/o}$ & $a\Gamma_\rho^\mathrm{w/}$ &
    $a\Gamma_\rho^\mathrm{av}$ &
    $\chi^2_\mathrm{w/o}$ & $\chi^2_\mathrm{w/}$ \\
    \hline\hline
    A30.32  & 0.12392(13) & 1.0081(52) & 0.0435(23) & 0.0427(30) & 0.0432(19) & 2.66 & 2.79 \\ 
    A40.24  & 0.14154(12) & 1.0206(95) & 0.0312(14) & 0.0243(15) & 0.0279(36) & 1.77 & 1.43 \\ 
    A40.32  & 0.14429(20) & 1.0039(28) & 0.0287(15) & 0.0271(18) & 0.0280(14) & 1.81 & 1.49 \\ 
    A60.24  & 0.17314(19) & 1.0099(49) & 0.0133(07) & 0.0139(07) & 0.0136(06) & 2.53 & 1.11 \\ 
    A80.24  & 0.19909(17) & 1.0057(29) & 0.0036(03) & 0.0040(05) & 0.0037(03) & 1.72 & 0.54 \\ 
    A100.24 & 0.22236(23) & 1.0037(19) & 0.0003(01) & 0.0004(01) & 0.0004(01) & 0.41 & 8.14 \\ 
    \hline
    B25.32  & 0.10850(32) & 1.0136(60) & 0.0454(50) & 0.0427(89) & 0.0447(46) & 1.05 & 0.56 \\ 
    B35.32  & 0.12380(10) & 1.0069(32) & 0.0340(20) & 0.0260(26) & 0.0309(43) & 0.97 & 0.90 \\ 
    B35.48  & 0.12486(14) & -          & 0.0316(24) & 0.0397(56) & 0.0328(46) & 1.35 & 0.88 \\ 
    B55.32  & 0.15551(12) & 1.0027(14) & 0.0123(05) & 0.0156(07) & 0.0136(17) & 1.30 & 0.93 \\ 
    \hline                                                                          
    D15.48  & 0.07067(15) & 1.0081(22) & 0.0491(114) &         - & 0.0491(114)& 0.68 & -    \\ 
    D30.48  & 0.09754(14) & 1.0021(07) & 0.0179(25) & 0.0206(40) & 0.0187(25) & 1.03 & 2.79 \\ 
    D45.32  & 0.12046(19) & 1.0047(14) & 0.0102(06) & 0.0079(15) & 0.0098(13) & 1.17 & 0.93 \\ 
    \hline\hline
  \end{tabular*}
  \caption{We give $aM_\pi$, the finite size correction factor
    $K_{M_\pi}$, the $\rho$ width $a\Gamma_\rho$ computed from $aM_\rho$ and
    $g_{\rho\pi\pi}$ using \eq{eq:rho_width}
    w/ and w/o thermal state
    removal, and the weighted average as explained in the text. In
    addition we give the reduced $\chi^2$-values of the 
    corresponding fits to the phase shift data.}
  \label{tab:width}
\end{table*}
 
We have two groups of ensembles with all identical parameters apart from the
volume. These are ensembles A40.24 and A40.32 as well as B35.32 and
B35.48, which we can use to investigate residual finite volume effects in our
results for $M_\rho$ and $\Gamma_\rho$.

\begin{figure}[th]
  \centering
  \subfigure{\includegraphics[width=.48\textwidth]{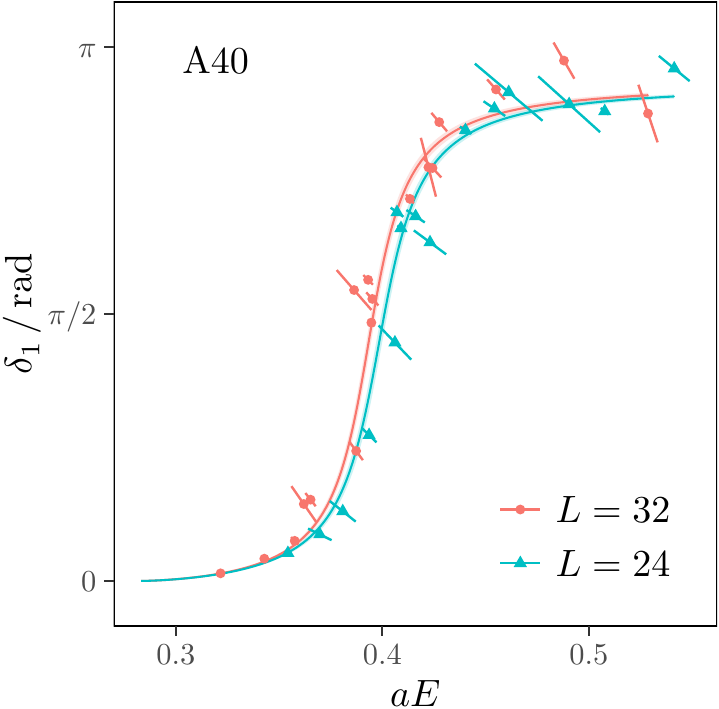}}
  \subfigure{\includegraphics[width=.48\textwidth]{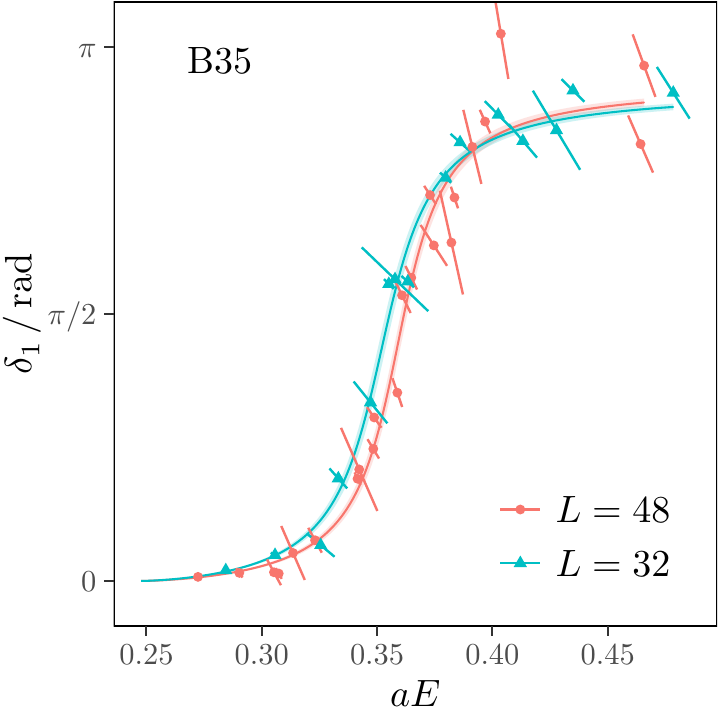}}
  \caption{We show the phase-shift $\delta_1$ as a function of
    $E_\mathrm{CM}$ in lattice units. Left we compare A40.24 (blue) with
    A40.32 (red) and right B35.48 (red) with B35.32 (blue). The lines
    with error bars represent the corresponding fits with
    \eq{eq:tan_delta} to the data.}
  \label{fig:phaseA40B35}
\end{figure}

In \fig{fig:phaseA40B35} we compare in the left panel the phase shift
points for A40.24 (blue) with the ones for A40.32 (red), in the right
panel B35.48 (red) with B35.32 (blue). Even though the Breit-Wigner fits
happen to result in slightly different values for the resonance
parameters, deviations are below the $2\sigma$ level and do not show a
systematic ordering with volume, see \tab{tab:rho_mass} and \tab{tab:width}. 

Thus, the weighted average with error including the systematic
uncertainty from thermal state removal should also safely include
residual effects from finite volume.

There are a few ensembles where the Breit-Wigner type fits to the
phase shift points are problematic. On the one hand this is the case
for ensemble with the heaviest pion mass A100.24. The
width approaches zero, which leaves the fits little freedom; a fact 
reflected by the untrustworthy $\chi^2$.

On the other hand, unfortunately the fit on D15.48, our most chiral
ensemble, is difficult, however, for different reasons. For D15.48
statistical uncertainties on the energy levels are quite large. As a
consequence, the Breit-Wigner fit for the case w/ thermal state
removal is not converging. The fit for the case w/o thermal state
removal gives large uncertainties. Combined with the rather low
lying inelastic threshold at $2M_K$, we do not consider this ensemble as
trustworthy for this calculation. 

\subsection{Chiral extrapolation}

\begin{figure}[t]
  \centering
  \subfigure{\includegraphics[width=.48\textwidth]{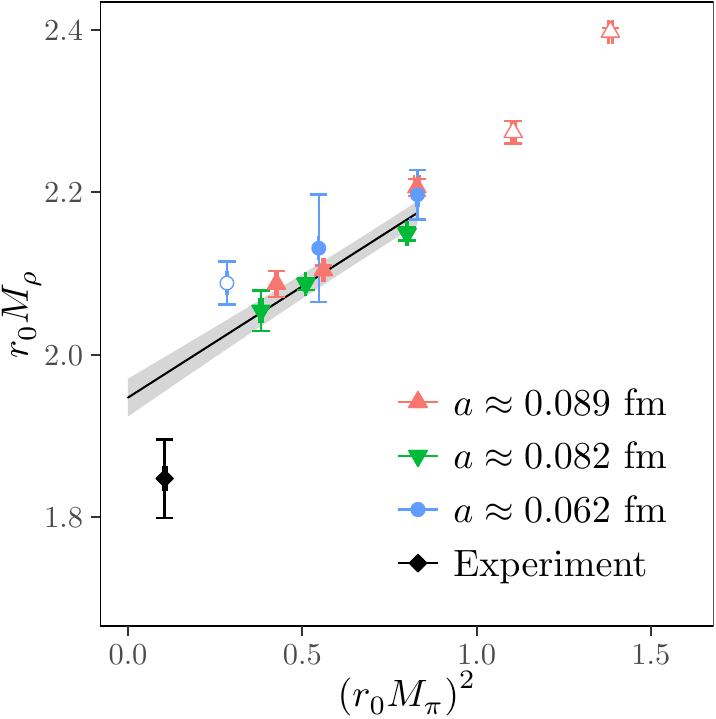}}
  \subfigure{\includegraphics[width=.48\textwidth]{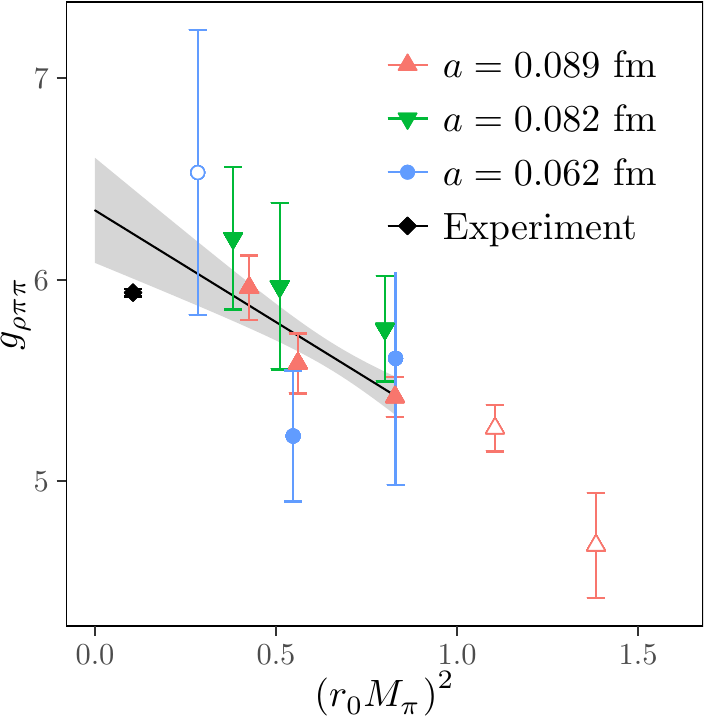}}
  \caption{In the left panel we show $r_0M_\rho^\mathrm{av}$ as a
    function of $(r_0M_\pi)^2$. Open symbols are not included in the
    fit. In the right panel $g_{\rho\pi\pi}^\mathrm{av}$ is shown also
    as a function of $(r_0M_\pi)^2$. The lines with error bands
    represent independent fits to the data.}
  \label{fig:Mrhog}
\end{figure}

We first consider $M_\rho$ and $g_{\rho\pi\pi}$. In the left panel of
\fig{fig:Mrhog} we show $r_0M_\rho^\mathrm{av}$, in the right
one $g_{\rho\pi\pi}^\mathrm{av}$, both as a function of
$(r_0M_\pi)^2$. Note that the error on $r_0/a$ is not included in the 
plot, because it is 100\% correlated for all data points of the same
$\beta$-value. Colours and symbols encode the three lattice spacing
values. The black diamonds represent the corresponding experimental values. The first
observation is that lattice artefacts are not resolvable given our
current level of statistical uncertainty. Overall, $M_\rho$ appears to
show a rather linear dependence on $M_\pi^2$, a bit less so
$g_{\rho\pi\pi}$. The values for $aM_\pi$ can be found in
\tab{tab:width}. For the following extrapolations we correct $aM_\pi$
for finite size effects by applying a correction factor $K_{M_\pi}$ computed in
Ref.~\cite{Carrasco:2014cwa}, which can also be found in
\tab{tab:width}.

Next we have tried to fit the pion mass dependence of
$M_\rho^\mathrm{av}$ and $g_{\rho\pi\pi}^\mathrm{av}$ combining 
Eqs.~(\ref{eq:brunsmeissner}) and (\ref{eq:grpipi}) up to the order
$M_\pi^3$. However, such a fit did not result in convincing
results. Even though the chiral log in $g_{\rho\pi\pi}^\mathrm{av}$
stemming from $f_\pi$ somewhat compensates the term $c_1M_\pi^2$, a
satisfactory description of the data for both the mass and the coupling 
could not be achieved. 

Therefore, we show in \fig{fig:Mrhog} independent linear
extrapolations for both $M_\rho$ and $g_{\rho\pi\pi}$ in $M_\pi^2$. As
visible, the two extrapolations overestimate both the $\rho$ mass and
the coupling at the physical point compared to experiment. 

\begin{figure}
  \centering
  \subfigure{\includegraphics[width=.48\textwidth]{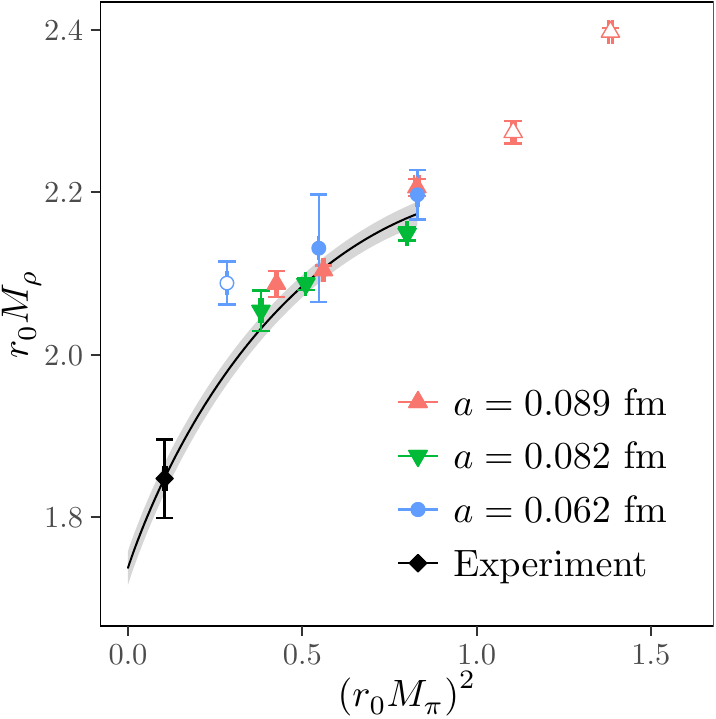}}
  \subfigure{\includegraphics[width=.48\textwidth]{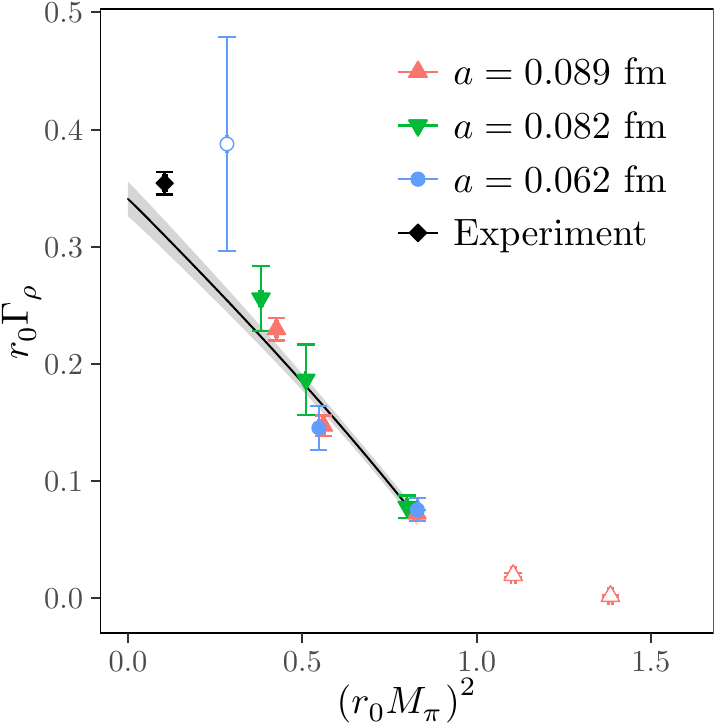}}
  \caption{Chiral extrapolation of $M_\rho$ and $\Gamma_\rho$ as a
    function of $M_\pi^2$, all in units of the Sommer parameter
    $r_0$. The lattice spacing is colour and symbol coded, the
    experimental  values are shown as black diamonds. The lines with
    error bands represent combined fits according to \eq{eq:fitform}
    to the data of $M_\rho$ and $\Gamma_\rho$. 
    Data points with
    open symbols are not included in the fit.}
  \label{fig:chiralextr}
\end{figure}

We now turn to combined fits of mass and width using
\eq{eq:feng_eq19_sommer} for the complex valued variable $Z$.
As described in section \ref{sec:method},
we extrapolate $M_\rho$ and $\Gamma_\rho$ to the physical point
combined in $r_0^2 Z=r_0^2(M_\rho + \mathrm{i}\Gamma_\rho/2)^2$. As we also
mentioned already, the error analysis for this fit is performed using
the parametric bootstrap procedure maintaining the correlation among
$M_\rho$, $\Gamma_\rho$ and $M_\pi$. We use $1500$
bootstrap samples and the values for $r_0/a$ for the different lattice
spacings were resampled from the values compiled in
\tab{tab:r0values}. 

The actual fit function reads
\begin{equation}
  \label{eq:fitform}
  \begin{split}
    a^2 Z &= p_{\sommer/a}^{-2} \left( 
    (p_1 + i p_2)
    + p_3 \left( p_{\sommer/a} a M_{\Pgp} \right)^2\right.\\
    &\quad- p_4 \sqrt{p_1 + i p_2} \left(p_{\sommer/a} a M_{\Pgp} \right)^3\\
    &\left.\quad+ (p_5 + i p_6) \ p_{\sommer/a}^{-2} 
    \right)\,.
  \end{split}
\end{equation}
The fit parameters are the following: $p_1$ and $p_2$ represent the
real and imaginary parts of $r_0^2Z_\chi$ and $p_3$ represents
$C_\chi$, furthermore $p_4 \equiv g^2_{\omega\rho\pi}/(24\pi r_0^2)$ and $p_5$
and $p_6$ parametrise the real and imaginary part of the $a^2$ lattice
artefacts. $p_{r_0/a}$ is one fit parameter per lattice spacing value
for $r_0/a$ accompanied by a corresponding prior $P_{r_0/a}$. Thus, we
have in total $6$ real-valued free fit parameters.

In the fit we include only the ensembles with the largest volume per
pion mass value, i.e. A40.24 and B35.32 are not included in the fit.
We do not include ensemble D15.48 in the fit, for reasons mentioned
above. Moreover, we include only data points with $M_\pi\leq
420\ \mathrm{MeV}$, which excludes ensembles A80.24 and A100.24.

The best fit parameters can be found in
\tab{tab:feng_20111_eq19_chiral_fit_r0} together with the reduced
$\chi^2$-value. We give the best fit parameters for 
fits with and without lattice artefacts included. Clearly, $p_5$ and
$p_6$, which parametrise the $a^2$ effects in $Z$ are compatible with
zero. Also, the remaining parameters do not change significantly with
and without $a^2$ artefact included in the fit.

\begin{table}[ht]
  \centering
  \begin{tabular*}{.45\textwidth}{@{\extracolsep{\fill}}lrr}
    \hline\hline
    Parameter & incl. $a^2$ & excl. $a^2$ \\ 
    \hline\hline
    $p_1$ & 3.14(28) & 2.99(07) \\ 
    $p_2$ & -0.631(61) & -0.592(26) \\ 
    $p_3$ & 4.75(24) & 4.79(08) \\ 
    $p_4$ & 0.936(80) & 0.991(34) \\ 
    $p_5$ & -5(10) & - \\ 
    $p_6$ & 1.3(1.8) & - \\ 
    $\chi^2 / \text{d.o.f.}$ & 2.35 & 2.00 \\ 
    \hline
  \end{tabular*}
  \caption{Best fit parameters of the combined chiral fit in terms of
    $Z$ with and without lattice artefacts included in the
    fit.}
  \label{tab:feng_20111_eq19_chiral_fit_r0}
\end{table}
 
The $\chi^2$-values for these fits are all a bit too large, indicating
a tension in the data in particular between $M_\rho$ and
$\Gamma_\rho$. It basically is a consequence of the invisible
curvature in the data for $M_\rho$.

The result of the fit can be seen in \fig{fig:chiralextr}, where we
show in the left panel $r_0M_\rho$ and in the right panel $r_0\Gamma_\rho$
both as functions of $(r_0 M_\pi)^2$. Note that the error on $r_0/a$
is not included in the plot, because it is 100\% correlated for all
data points of the same $\beta$-value.
The best fit to the data is
indicated by the solid lines with error bands. Data points with open
symbols are excluded from the fit. The fit range is indicated by the
extent of the solid lines. The experimental values are included in both
plots as black diamonds, but not included in the fit. 

Our final result for $M_\rho$ and $\Gamma_\rho$ taken from the fit
without a $a^2$ effects included reads
\begin{equation}
  \label{eq:final}
  M_\rho\ =\ 769(19)\ \mathrm{MeV}\,,\qquad\Gamma_\rho\ =\ 129(7)\ \mathrm{MeV}\,,
\end{equation}
corresponding to
\begin{equation}
  g_{\rho\pi\pi}\ =\ 5.5(1)\,.
\end{equation}
In addition we find
\begin{equation}
  \label{eq:gorp}
  \begin{split}
    M_\rho^0\ &=\ 723(20)\ \mathrm{MeV}\,,\\
    \Gamma_\rho^0\ &=\ 142(7)\ \mathrm{MeV}\,,\\
    |g_{\omega\rho\pi}|\ &=\ 20.8(7)\ \mathrm{GeV}^{-1}\\
  \end{split}
\end{equation}
from our chiral and continuum fits. The correlation coefficients for
the fit parameters can be found in appendix~\ref{sec:appcor}.

\section{Discussion}
\label{sec:discussion}

\begin{figure*}[t]
  \centering
  \includegraphics[width=.85\textwidth]{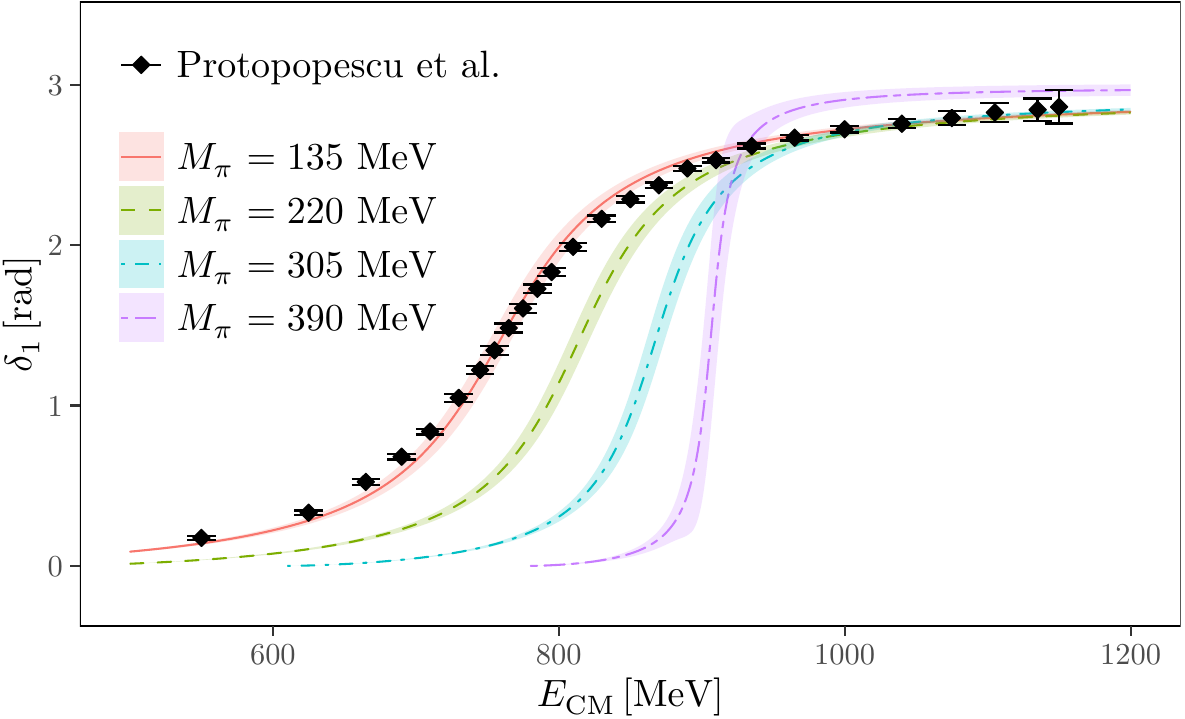}
  \caption{Comparison of experimental phase shift data from
    Ref.~\cite{Protopopescu:1973sh} to the phase shift curve
    extracted from our final results for $M_\rho$ and
    $\Gamma_\rho$ shown as red solid line. For illustration purposes we also 
    show the phase shift curve in a world with $M_\pi=220\ \mathrm{MeV}$ as green 
    dashed line, with $M_\pi=305\ \mathrm{MeV}$ as blue dot-dashed line and
    with $390\ \mathrm{MeV}$ as a purple two-dashed line.}
  \label{fig:chiralphase}
\end{figure*}

In the previous section we have performed different chiral and
continuum extrapolations for our data. First, there are independent
linear fits of $M_\rho$ and $g_{\rho\pi\pi}$ as a function of
$M_\pi^2$. Second, we have performed combined fits in terms of $Z$ as
function of $M_\pi$ including terms up to order $M_\pi^3$ with and
without including lattice artefacts. While the two linear fits
certainly provide a good description of the data for $M_\rho$ and
$g_{\rho\pi\pi}$ separately, we decided to quote the results from the
combined fit as our final result. The reason is that in the
corresponding effective field theory the complex pole is treated
consistently, which we consider as theoretically more sound.

The final result for $M_\rho$ and $\Gamma_\rho$ we quote in
\eq{eq:final} can be compared to the corresponding PDG
values~\cite{Tanabashi:2018oca} for mass and full width
\[
M_\rho^\mathrm{exp}\ =\ 775.26(25)\ \mathrm{MeV}\,,\qquad
\Gamma_\rho^\mathrm{exp}\ =\ 149.1(8)\ \mathrm{MeV}\,.
\]
Note that these also correspond to Breit-Wigner parameters determined
experimentally from $e^+e^-$ reactions. The deviation to other
reactions can be of the order of $10\ \mathrm{MeV}$.
We observe rather good agreement for $M_\rho$, while our value for the
width is slightly too low. This is also visible in \fig{fig:chiralphase},
where we plot the experimental phase shifts from
Ref.~\cite{Protopopescu:1973sh} and 
compare them to the phase shift curve we obtain using the final values from
\eq{eq:final} and then again assuming the Breit-Wigner form from \eq{eq:tan_delta}.

However, this good agreement should be taken with caution. First of
all our extrapolation form for $M_\rho$ and $\Gamma_\rho$ is not model
independent. This is in particular important, because the curvature
needed to obtain an $M_\rho$-value close to the experimental one comes
from constrains due to $\Gamma_\rho$. This, as discussed earlier,
manifests itself also in a bit too large $\chi^2$-values in the chiral
and continuum fits. Moreover, the ensemble with the
lightest pion mass included in the fit is B35.48 with a pion mass of
about $300\ \mathrm{MeV}$. Thus, the extrapolation to the physical
point is quite long. In addition we have assumed that we can perform a
Breit-Wigner type fit to all the phase shift data, which is an
approximation. This might also be the reason for the too low value of
$\Gamma_\rho$ compared to experiment. We are currently working on an
alternative extrapolation using the inverse amplitude method which
might allow us to perform the chiral extrapolation even more
reliably~\cite{Truong:1991gv,Dobado:1992ha,Dobado:1996ps,GomezNicola:2007qj,Niehus:2019nkl}. 
Our fitted value for $g_{\omega\rho\pi}$ \eq{eq:gorp} is in the right
ballpark, when compared to the numbers given in
Refs.~\cite{Djukanovic:2009zn,Djukanovic:2010id}, where
$16\ \mathrm{GeV}^{-1}$ is quoted. From
Refs.~\cite{Meissner:1987ge,Kaiser:1990yf} one finds
$g_{\omega\rho\pi}=\pm20.7\ \mathrm{GeV}^{-1}$ in very good agreement
with our value.

Finally, our determinations of mass and width rest on the assumption
that all partial waves apart from $\ell=1$ are negligible. This
assumption is supported by previous lattice investigations of the
$\rho$ meson, but has not been checked by us yet.

On the other hand, our results for $M_\rho$ and $\Gamma_\rho$ make a 
combined extrapolation to the physical point and to the continuum limit 
feasible for the first time. However, since we find lattice artefacts to be
statistically insignificant, our final result is based on a chiral
extrapolation assuming no lattice artefacts. We have different
volumes available with otherwise fixed parameters, which allow us to
argue that residual finite volume effects are not a dominant source of
uncertainty in our results. 

\begin{figure*}
  \centering
  \includegraphics[width=\textwidth]{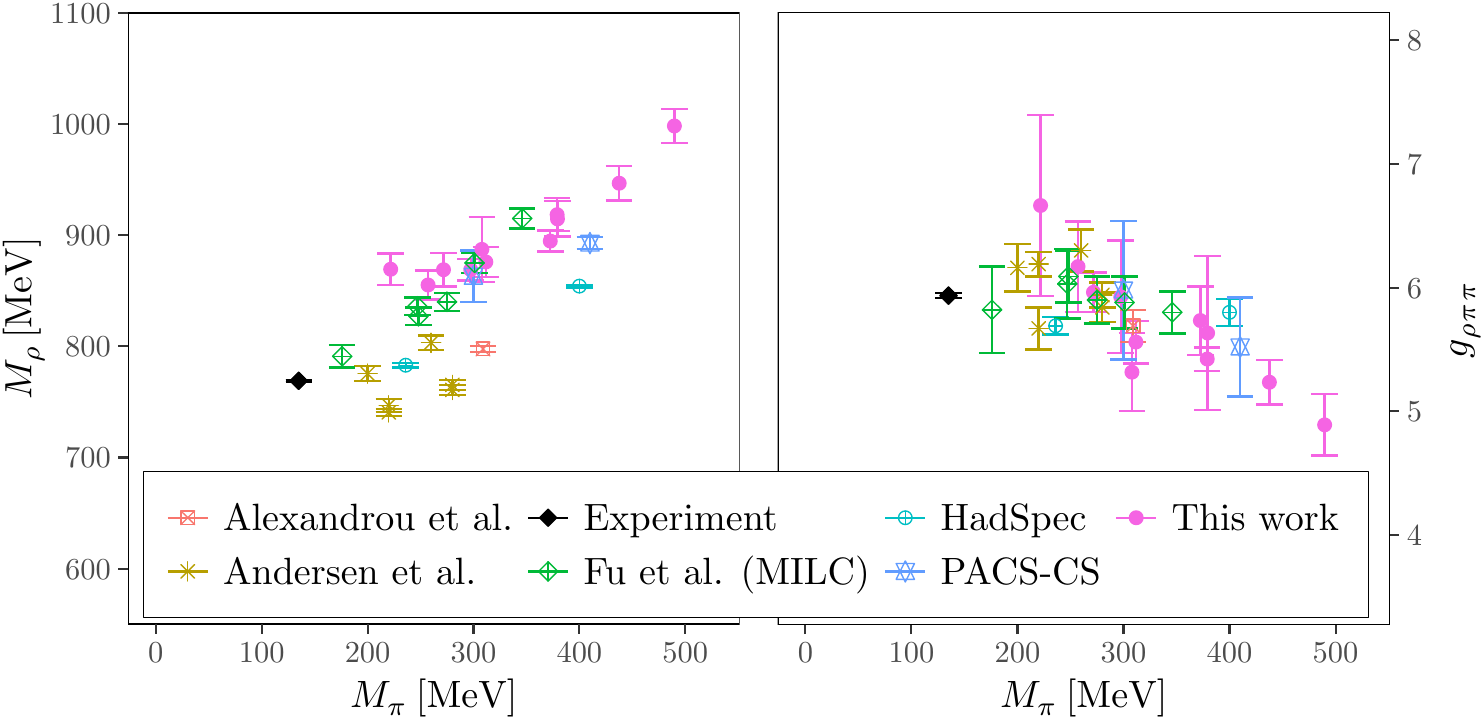}
  \caption{Comparison of lattice results for $M_\rho$ (left) and
    $g_{\rho\pi\pi}$ (right) as a function of $M_\pi$. We compare with all 
    available results that had a dynamic strange quark: 
    Alexandrou et \textit{al.}~\cite{Alexandrou:2017mpi}, 
    Andersen et \textit{al.}~\cite{Andersen:2018mau},
    Fu et \textit{al}.~\cite{Fu:2016itp},
    HadSpec~\cite{Dudek:2012xn,Wilson:2015dqa},
    PACS-CS~\cite{Aoki:2011yj} as well as the experimental
    value~\cite{Tanabashi:2018oca}.} 
  \label{fig:LatComp}
\end{figure*}

In \fig{fig:LatComp} we compare results for $M_\rho$ and
$g_{\rho\pi\pi}$ from various lattice collaborations with $N_f=2+1$ or
$N_f=2+1+1$ dynamical quark flavours. We observe that there are
probably lattice artefacts in some of the results for $M_\rho$, in
particular in the results from Andersen et
\textit{al.}~\cite{Andersen:2018mau} and from the Hadron Spectrum
Collaboration~\cite{Dudek:2012xn}. For $g_{\rho\pi\pi}$ uncertainties 
are in general larger and within these large uncertainties the
agreement among different lattice collaborations is reasonable. 

However, leaving aside lattice artefacts, one could be tempted to
conclude from \fig{fig:LatComp} that $M_\rho$ is rather linear in
$M_\pi$, very similar to what is observed for the nucleon
mass~\cite{Walker-Loud:2014iea}. In fact one finds $M_\rho = 680\ \mathrm{MeV}
+ 0.6 M_\pi$ to a good approximation by fitting only the data by Fu and
Wang~\cite{Fu:2016itp} together with our data, which represents yet another
version of the \enquote{ruler} plot. From an effective field theory
point of view this cannot be the correct pion
mass dependence and future results will hopefully shed light on this puzzle.

We can also compare to the results of Ref.~\cite{Giusti:2018mdh},
where $M_\rho$ and $g_{\rho\pi\pi}$ have been determined on the same
ETMC ensembles we used, however, using the inverse Lüscher
method based only on the vector current based on a parametrisation of
the pion form factor. Their continuum extrapolated values for $M_\rho$
and $g_{\rho\pi\pi}$ at the physical point are consistent with ours.

\section{Summary}
\label{sec:summary}

We have presented an investigation of the $\rho$-meson properties
using lattice QCD with $N_f=2+1+1$ Wilson twisted mass quarks at
maximal twist. With three values of the lattice spacing and a range of
pion mass values we could perform chiral and continuum extrapolations
of $\rho$-meson mass $M_\rho$ and width $\Gamma_\rho$ with better
control than previously possible. The latter two quantities have been
determined on our ensembles using a Breit-Wigner type fit to phase
shift data assuming that partial waves with $\ell\geq3$ are negligible.

The phase shift curves have been determined applying Lüscher's method
using moving frames up to $\bm{d}^2=4$ and all available lattice
irreducible representations. Our final result reads
\[
\begin{split}
  M_\rho\ &=\ 769(19)\ \mathrm{MeV}\,,\\
  g_{\rho\pi\pi}\ &=\ 5.5(1)\,,\\
  \Gamma_\rho\ &=\ 129(7)\ \mathrm{MeV}\,,\\
\end{split}
\]
which is determined from a combined continuum and chiral extrapolation
of $M_\rho$ and $\Gamma_\rho$.
Systematic errors from thermal state pollutions, the chiral and
the continuum extrapolation should be covered by the error we
quote. $M_\rho$ is very close to its experimental value, the width is
about two sigma too low. The agreement of our data for $M_\rho$ with
previously published lattice results is satisfactory. 

It is clear that more work is needed to better estimate the width,
which likely suffers from e.g. the use of a Breit-Wigner type fit to
the phase shift data. Therefore, we are currently working on using the
inverse amplitude method to directly describe the pion mass dependence
of the phase shift curves~\cite{Niehus:2019nkl}, see also
Ref.~\cite{Hu:2017wli}. This should alleviate systematic uncertainties
in our current analysis.

\section*{Acknowledgements}

We thank the
members of ETMC for the most enjoyable collaboration.
We thank X.~Feng, B.~Kubis, U.-G.~Meißner and A.~Rusetsky for very
useful discussions and valuable comments to the draft. We thank two
anonymous referees for helpful comments.
The authors gratefully acknowledge the Gauss Centre for Supercomputing
e.V. (www.gauss-centre.eu) for funding this project by providing
computing time on the GCS Supercomputer JUQUEEN~\cite{juqueen} and the
John von Neumann Institute for Computing (NIC) for computing time
provided on the supercomputer JURECA~\cite{jureca} at Jülich
Supercomputing Centre (JSC).
This project was funded by the DFG as a project 
in the Sino-German CRC110. The open source software
packages tmLQCD~\cite{Jansen:2009xp,Abdel-Rehim:2013wba,Deuzeman:2013xaa}, Lemon~\cite{Deuzeman:2011wz},
QUDA~\cite{Clark:2009wm,Babich:2011np,Clark:2016rdz} and 
R~\cite{R:2005} have been used.

\bibliographystyle{h-physrev5}

\begin{appendix}
   \section{Operator construction}
\label{sec:operator_construction}

One side effect of the restriction to finite volumes in a lattice
calculation is, that the symmetry group of rotoflections (rotations
and space inversions) is reduced to a finite subset. 
In sections~\ref{sec:luescher} and \ref{sec:operators}, the
consequences of this explicit symmetry breaking were encapsulated in a
set of subduction coefficients. Here we illustrate our derivation of
subduction coefficients as well as the chosen conventions for a single
and two pions.

Let $\ket{l,m}$ be a basis vector in the spherical basis transforming according to the angular momentum-$l$
representation of $\SO(3)$ and $m$ denote the magnetic quantum number. For a given rotation $R$ and basis vectors $\ket{l, m}$, the representation matrix elements are given by 
\begin{align}
  D^l_{m,m'}(R) &= \mel{l,m}{\R}{l,m'} \,,
  \label{eq:rotation_operator_matrix_elements}
\end{align}
where \R denotes the action of $R$ on the Hilbert space of wave functions.

Let $\G$ be the (finite) symmetry group of the discretised geometry. As already explained in \eq{eq:irrep_demomposition}, $D^l$ is not necessarily irreducible over $\G$ and it may decompose into multiple irreducible representations $\Gamma$ of $\G$. Then
\begin{align}
  \hat{P}_{\alpha \beta}^{\Gamma, l}
  &= \frac{\mathrm{dim}(\Gamma)}{\abs{\G}}
  \sum\limits_{g \in \G} D^\Gamma(R_g)^*_{\alpha \beta} \R_g
  \label{eq:projection_subgroup_general}
\end{align}
defines a projector, where $D^\Gamma$ denote the irreducible
representation matrices and  
$\alpha, \beta \,\in \,\left\{ 1,\ldots,\mathrm{dim}(\Gamma) \right\}$ are 
arbitrary but fixed. We refrain from discussing the modifications for
non-trivial multiplicities. We use the Schönflies notation and
follow the conventions for $D^\Gamma$ used in crystallography
\cite{altmann1994point}, conveniently implemented in Maple
\cite{rykhlinskaya2005generation}. 

\begin{table}
	\centering
	\begin{tabular}{lcc}
		\hline\hline
    $\bm{d}^2$ & $\LG(\pcm)$ & $\Gamma$ \\
		\hline\hline
    $0$ & $\Oh$      & $\mathrm{T{1u}}$ \\
    $1$ & $\Cfourv$  & $\mathrm{A1} \oplus \mathrm{E}$ \\
    $2$ & $\Ctwov$   & $\mathrm{A1} \oplus \mathrm{B1} \oplus \mathrm{B2}$ \\ 
		$3$ & $\Cthreev$ & $\mathrm{A1} \oplus \mathrm{E}$ \\
		$4$ & $\Cfourv$  & $\mathrm{A1} \oplus \mathrm{E}$ \\

		\hline\hline
	\end{tabular}
	\caption{Little groups and decomposition of angular momentum 1 for all momentum sectors 
    $\bm{d}^2$ used in this work. The groups are isomorphic for each
    representative of a sector. Therefore, the direction of $\pcm$ is arbitrary here.} 
	\label{tab:little_groups}
\end{table}

At rest, the symmetry group is the octahedral group $\Oh$. Choosing a
non-zero CM-momentum $\pcm$ further reduces the relevant symmetry
group to the \enquote{little group}  
\begin{equation}
  \LG(\pcm) \equiv \set{g \in \Oh, \R_g \ \pcm = \pcm}\,,
\end{equation}
which leaves $\pcm$ invariant. The relevant little groups are listed
in \tab{tab:little_groups}. 

\subsection{One-meson operator}

Let $\mathcal{O}_l^{m \dagger}(\pcm)$ be an operator that creates a
meson state $\ket{\pcm; l,m}$ with momentum $\pcm$ and total (integral)
angular momentum $l$ with projection $m$.
By applying $\hat{P}^{\Gamma, l}$, this operator is projected to an operator
\begin{align}
  \mathcal{O}_\Gamma^{\alpha \dagger}(\pcm)
  &= \sum_{\beta} \phi_\beta \sum_{m} \phi_{m} \hat{P}_{\alpha \beta}^{\Gamma, l}\,
  \mathcal{O}_l^{m\dagger}(\pcm) \nonumber \\
  &=  \sum_{\beta} \phi_\beta \sum\limits_{m,m'} \phi_{m}
  \frac{\mathrm{dim}(\Gamma)}{\abs{\G}} \nonumber \\
  &\times        \sum\limits_{g \in \LG(\pcm)} D^\Gamma(R_g)^*_{\alpha \beta}\,D^l_{m',m}(R_g)
  \mathcal{O}_l^{m' \dagger}(\pcm) \,,
  \label{eq:projected_basis}
\end{align}
which creates a single meson basis state $\ket{\pcm; \Gamma, \alpha}$ of $\LG(\pcm)$. 

Here it becomes apparent, why $\alpha$ are called \enquote{rows}.
The row index of the matrix $D^\Gamma$ also labels the basis vectors of
$\Gamma$. Correspondingly we will refer to $\beta$ as the
\enquote{column} of the representation. $\phi_m$ and $\phi_\beta$ are
phases which are chosen such that the set $\ket{\pcm; \Gamma, \alpha}$ become
orthonormal. In the following we suppress the dependence on $\beta$
and $\phi$. We denote the coefficients with fixed phases by the
\enquote{subduction coefficient} $s^{\Gamma}_{l}$ and from
\eq{eq:projected_basis} obtain the result
\begin{align}
  \mathcal{O}_\Gamma^{\alpha \dagger}(\pcm) = \sum\limits_{m'} s^{\Gamma, \alpha}_{l, m'}\, \mathcal{O}_l^{m' \dagger}(\pcm) \,.
  \label{eq:one_meson_operator}
\end{align}
Applying the creation operators on the left and right side to a vacuum state yields
\eq{eq:subduced_basis} for the subduction of basis states. Note that
the projection only acts in the space of total angular momentum. The
linear momentum $\bm{p}$ is unaffected by the procedure.

\subsection{Two-pion operators}

\begin{table*}
  \centering
  \begin{tabular}{llc}
    \hline\hline
    $\bm{d}$ & $\Gamma$  &     $\bm{p}_1 \otimes \bm{p}_2$ \\
    \hline\hline
    $(0, 0, 0)$ &     $\mathrm{T1u}$ &  $(0, 0, 1) \otimes (0, 0, -1)$, $(1, 0, 1) \otimes (-1, 0, -1)$ \\
    
    $(0, 0, 1)$ &     $\mathrm{A1}$ &   $(0, 0, 1) \otimes (0, 0, 0)$, $(0, 0, 2) \otimes (0, 0, -1)$, $(1, 0, 1) \otimes (-1, 0, 0)$, $(1, 1, 1) \otimes (-1, -1, 0)$ \\
    
    $(0, 0, 1)$ &     $\mathrm{E}$ &    $(0, 1, 1) \otimes (0, -1, 0)$, $(1, 1, 1) \otimes (-1, -1, 0)$ \\
    
    $(1, 1, 0)$ &     $\mathrm{A1}$ &   $(1, 1, 0) \otimes (0, 0, 0)$, $(1, 1, 1) \otimes (0, 0, -1)$, $(1, -1, 0) \otimes  (0, 2, 0)$ \\
    
    $(1, 1, 0)$ &     $\mathrm{B1}$ &   $(1, 1, 1) \otimes (0, 0, -1)$, $(1, 0, 1) \otimes (0, 1, -1)$ \\
    
    $(1, 1, 0)$ &     $\mathrm{B2}$ &   $(1, 0, 0) \otimes (0, 1, 0)$, $(1, 0, 1) \otimes (0, 1, -1)$, $(2, 0, 0) \otimes (-1, 1, 0)$ \\
    
    $(1, 1, 1)$ &     $\mathrm{A1}$ &   $(1, 1, 1) \otimes (0, 0, 0)$, $(1, 0, 1) \otimes  (0, 1, 0)$, $(2, 0, 0) \otimes (-1, 1, 1)$ \\
    
    $(1, 1, 1)$ &     $\mathrm{E}$ &    $(1, 0, 1) \otimes (0, 1, 0)$, $(1, -1, 1) \otimes (0, 2, 0)$ \\
    
    $(0, 0, 2)$ &     $\mathrm{A1}$ &   $(0, 0, 2) \otimes (0, 0, 0)$ \\
    
    $(0, 0, 2)$ &     $\mathrm{E}$ &    $(0, 1, 1) \otimes (0, -1, 1)$ \\
    \hline\hline
  \end{tabular}
  \caption{Momentum combinations $\bm{p}_1 \otimes \bm{p}_2$ used in
    \eq{eq:two_pion_operator}.
    We only give one representative
    $\text{CM}$ momentum $\pcm = 2 \pi \bm{d} / L$ for each
    momentum sector. The other directions may be generated by a
    global rotation. The momentum combinations depend on the irrep
    $\Gamma$ because not all combinations couple to all irreps.}
  \label{tab:two_pion_list_of_q}
\end{table*}

To subduce the two-pion operators with individual 3-momenta
$\bm{p}_1,\,\bm{p}_2$ and $\pcm = \bm{p}_1 +\bm{p}_2$ into the
irreducible representations of the residual lattice rotation symmetry
group $\LG(\pcm)$ we start from the product operator
$\Pgpp(\bm{x}_1)\,\Pgpm(\bm{x}_2)$. Then our group projection formula
reads~\cite{Feng:2010es} 
\begin{equation}
  \begin{split}
    {\mathcal{O}_{\Pgp\Pgp}}^{\alpha \dagger}_{\Gamma \bm{q}}(\pcm)
    \quad &= \frac{\dim\left( \Gamma \right)}{|\LG(\pcm)|}\\
    &\times\sum\limits_{\beta} \phi_\beta \,
     \sum_{g \in \LG(\pcm)} \sum_{\bm{x}_1, \bm{x}_2}\\
    &\times\mathrm{e}^{i (\bm{x}_1 \cdot (
      \frac{1}{2} \pcm + \R_g  \bm{q}) + \bm{x}_2 \cdot (\frac{1}{2} \pcm - \R_g \bm{q}))}  \,\\
    &\times D^\Gamma(R_g)_{\alpha \beta}^*  \,
    \mathcal{O}^\dagger_{\Pgpp}(\bm{x}_1) \, \mathcal{O}^\dagger_{\Pgpm}(\bm{x}_2) \,,\\
  \end{split} 
  \label{eq:two_pion_operator}
\end{equation}
where $2\bm{q} = \bm{p}_1 - \bm{p}_2 $ and $\alpha = 1,\cdots,\dim\left( \Gamma \right)$.
The vector $\phi = \left( \phi_1,\cdots,\phi_{\dim\left( \Gamma \right)} \right)$ characterises
again our choice of phase and normalisation for the irreducible operator multiplet. 

Two-pion operators in the same reference frame $\pcm$ but with
different relative momenta $\bm{q} \ne \bm{q}'$, which are related by
an element of $\LG(\pcm)$, $\R_g \bm{q} = \bm{q'}$ for some  $g \in
\LG(\pcm)$, lead to linearly dependent operators under  the projection
Eq.~\ref{eq:two_pion_operator}. Therefore, we only use certain
momentum combinations $\bm{p}_1 \otimes \bm{p}_2$. In
Tab.~\ref{tab:two_pion_list_of_q} we list one representative
combination for each momentum sector. The two-pion operators for
unlisted moving frames $\pcm'$ with $|\pcm| = |\pcm'|$ are constructed
by a global rotation for which $\R_{\tilde{g}} \pcm = \pcm'$.

The method we describe here can be understood as an extension of the
projection method of Ref.~\cite{Prelovsek:2016iyo} for arbitrary
moving frames.

\section{Analysis Details}
\label{sec:details}

In this appendix we give the details on our analysis to estimate the
extrapolated values for $M_\rho$ and $\Gamma_\rho$ starting from
energy levels $aE_\mathrm{CM}$ and $aM_\pi$.

On a per ensemble basis we use the various interacting energy levels
$aE_\mathrm{CM}$ together with the values of $aM_\pi$ to determine
phase shift values $\delta_1(E_\mathrm{CM})$. For the reasons
explained above we use in this step the jack-knife procedure to
estimate the variance-covariance matrix for all $aE_\mathrm{CM}$,
$\delta_1$ and $M_\pi$ using the standard jack-knife estimators. In
particular, the Lüscher function is evaluated on the jack-knife
samples. 

Since with jack-knife there are not neccessarily identical numbers of
replicates for all ensembles, we now use parametric bootstrap to
resample the distributions with $1500$ bootstrap
replicates on each ensemble. With all the mean values and the
variance-covariance matrix as input we draw random samples from a
corresponding multivariate Gaussian distribution. The such generated
bootstrap replicates fully reproduce the input variance-covariance
matrix.

Generating multi-variate Gaussian random variables $Y$ with a given
symmetric, positive definite covariance matrix $C$ from independent
standard normal random variables $X$ can be performed as
follows: 
\[
\begin{split}
  Y = \sqrt{C} X\quad&\Rightarrow\\
  \quad \mathrm{Cov}(Y,
  Y)\ &=\ \langle Y\cdot Y^t\rangle = 
  \sqrt{C}\, \langle X \cdot X^t\,\rangle\, \sqrt{C}^t\ =\ C\,,\\
\end{split}
\]
since $\langle X\cdot X^t\rangle=1$.

In the next step the Breit-Wigner functional form is fitted to the
phase shift data for each ensemble seperately. Note that we could have
also performed these fits on the jack-knife samples and resample
afterwards. We actually did it both ways and found full agreement.

The fit including errors on the $x$-axis and including priors for fit
parameters is performed as follows (see also Ref.~\cite{hadron} for an
implementation): 
lets assume the proposed
functional form of the model reads
\[
y(x)\ =\ f(x, \alpha_1, \ldots, \alpha_{n_\alpha}; \beta_1, \ldots, \beta_{n_\beta})\,,
\]
which, for simplicity, we assume to be a scalar function.
Assume further that we have $n_d$ data points $y_1, \ldots, y_{n_d}$
at $x$-values $x_1, \ldots, x_{n_d}$ for all of which we have
estimates $\bar y_i$ and $\bar x_i$.
Moreover, we have estimates for the
$n_\alpha$ parameters $\alpha_i$ reading $\bar\alpha_i$. The remaining
parameters $\beta_j$ are free fit 
parameters. Then we may define the following function for fixed
$\beta=(\beta_1, \ldots,\beta_{n_\beta})$ 
\[
F: \mathbb{R}^{n}\ \to\ \mathbb{R}^{n+n_d}\,,\qquad Y\ = F(X; \beta)\,,
\]
with $n= n_d + n_\alpha$. The elements of $F$ are defined as follows
\[
F_i(X, \beta) =
\begin{cases}
  f(X_i, X_{n_d+1}, \ldots, X_n; \beta) &  1\leq i\leq n_d\,,\\
  X_{i-n_d} &  n_d < i \leq n + n_d \,. 
\end{cases}
\]
$X\in\mathbb{R}^{n}$ represents the concatenation of all the $x_i$
and all parameters $\alpha_i$ reading $X = (x_1, \ldots, x_{n_d},
\alpha_1, \ldots, \alpha_{n_\alpha})$. We perform a similar concatenation for
the data, i.e. $\bar y\in\mathbb{R}^{n+n_d}$ with $\bar y=(\bar y_1, \ldots, \bar
y_{n_d}, \bar x_1, \ldots, \bar x_{n_d}, \bar\alpha_1, \ldots,
\bar\alpha_{n_\alpha})$. Then one has to minimise
\[
\chi^2\ =\ (\bar y - F(X; \beta))\cdot C^{-1}\cdot (\bar y - F(X; \beta))^t
\]
over $X$ and $\beta$ with $C = \langle Y\cdot Y^t\rangle$ the
variance-covariance matrix. $C$ is conveniently replaced by its
estimate $\bar C$ obtained from the corresponding jack-knife
estimator. We use the frozen variance-covariance matrix approximation,
where $C$ is kept fixed during the resampling.

In our case the parameters $\alpha_i$ correspond for instance to
$r_0/a$ at the different $\beta$-values or $\overline{M}_{\pi^+}$ used as
input. Of course, depending on the 
problem $C$ and $\bar C$ factorise into block diagonal form.

\section{Correlation Coefficients}\label{sec:appcor}

In the following table we compile the correlation coefficients of the
chiral fit without lattice artefacts included in the fit: fit function
is thus \eq{eq:fitform} without the term proportional to $(a/r_0)^2$.
The bare data can be made available upon request.

\begin{table*}[ht]
  \centering
  \begin{tabular}{lrrrrrrr}
    \hline
    & $p_1$ & $p_2$ & $p_3$ & $p_4$ & $p_{r_0/a}(A)$ & $p_{r_0/a}(B)$ & $p_{r_0/a}(D)$ \\
    \hline
    $p_1$ & 1.00 & -0.24 & -0.38 & -0.42 & 0.67 & 0.61 & 0.30 \\
    $p_2$ & -0.24 & 1.00 & -0.61 & -0.35 & -0.37 & -0.36 & -0.16 \\ 
    $p_3$ & -0.38 & -0.61 & 1.00 & 0.66 & 0.01 & 0.07 & -0.07 \\
    $p_4$ & -0.42 & -0.35 & 0.66 & 1.00 & -0.57 & -0.53 & -0.42 \\
    $p_{r_0/a}(A)$ & 0.67 & -0.37 & 0.01 & -0.57 & 1.00 & 0.87 & 0.50 \\
    $p_{r_0/a}(B)$ & 0.61 & -0.36 & 0.07 & -0.53 & 0.87 & 1.00 & 0.48 \\
    $p_{r_0/a}(D)$ & 0.30 & -0.16 & -0.07 & -0.42 & 0.50 & 0.48 & 1.00 \\
    \hline
  \end{tabular}
  \caption{Correlation coefficients of fit parameters corresponding to
  the chiral fit of \eq{eq:fitform} to our data.}
  \label{tab:feng_20111_eq19_chiral_fit_cormatrix}
\end{table*}

 \end{appendix}
\end{document}